\documentclass[epj,referee]{svjour}

\usepackage{graphicx}

\setkeys{Gin}{width=0.4\textwidth}

\begin{document}

\title{Small Bipolarons in the 2-dimensional Holstein-Hubbard Model}
\subtitle{I  The Adiabatic Limit}
\author{L. Proville  \thanks{\emph{Present Address:} DAMTP Cambridge
University, Cambridge, CB3 9EW, UK} \and S. Aubry}
\institute{Laboratoire L\'eon Brillouin (CEA-CNRS), CEA Saclay
	91191-Gif-sur-Yvette Cedex, France}

\date{Received: December 10 1998 / Revised: \today}
\abstract{The spatially localized bound states of two electrons in
	the adiabatic two-dimensional Holstein-Hubbard model on
	a square lattice are investigated both numerically and
	analytically. The interplay between the electron-phonon
	coupling $g$, which tends to form bipolarons and the
	repulsive Hubbard interaction $\upsilon \geq 0$, which tends to
	break them, generates many different ground-states.
	There are four domains in the $g,\upsilon$ phase diagram
	delimited by first order transition lines. Except for the domain
	at weak electron-phonon coupling (small $g$) where the
	electrons remain free, the electrons form bipolarons
	which can 1) be mostly located on a single site (small $\upsilon$,
	large $g$); 2) be an anisotropic pair of polarons
	lying on two neighboring sites in the magnetic singlet
	state (large $\upsilon$, large $g$); or 3) be a 
	"quadrisinglet state" which is the superposition of 4 electronic singlets
	with a common central site. This quadrisinglet bipolaron is 
	the most stable in a small central domain in between the three
	other phases.
	The pinning modes and the Peierls-Nabarro barrier of each of
	these bipolarons are calculated and the barrier is 
	found to be strongly depressed
	in the region of stability of the quadrisinglet bipolaron.
\PACS{
      {71.10.Fd}{Lattice fermion models (Hubbard model, etc.)}   \and
      {71.38.+i}{Polarons and electron phonon interactions} \and
      {74.20.Mn} {Nonconventional mechanisms (spin fluctuations, polarons and bipolarons,
      resonating valence bond model, anyon mechanism, marginal Fermi liquid, Luttinger liquid,
      etc.)} \and
      {74.25.Jb} {Electronic structure}}}

\maketitle

\section{Introduction} \label{intro}

The standard BCS theory of superconductivity \cite{BCS57} holds
for a system of noninteracting electrons weakly
coupled to a quantum field of phonons. It has been well-known for several
decades that when the electron-phonon coupling increases too much,
the BCS theory breaks down because of lattice instabilities \cite{Mig58}.
As a consequence, rather low critical temperatures ($\approx 30  K$)
were predicted as the upper bound for real BCS superconductors
\cite{McM68}. Many theories have subsequently been developed to
describe the strong coupling regime with the hope to predict the
existence of non-BCS superconductors with high critical temperature.
After the discovery by Bednorz and M\"uller \cite{BM86,MB93,Wal96}
of cuprate materials, which can be superconducting at temperatures
as high as $100 K$ or more, the bipolaron approach
(among others) regained much interest \cite{ASL95}.

Since Landau \cite{Lan33}, it has been acknowledged that a single
electron (or equivalently  a pair of noninteracting electrons
coupled to a deformable classical field) may localize in the potential
created self-consistently by a deformation of the field. The
resulting object is called "polaron" for one electron or "bipolaron"
for two electrons. The bipolaron theory of Alexandrov et al.\cite{AAR86} 
involves small bipolarons which are pairs of electrons with opposite spins,
sharply localized at single sites of the lattice. Actually, because the
phonons are quantum, these bipolarons are hard-core bosons that could
condense in a superfluid state.
For models in two dimensions and more, bipolarons exist only
when the electron-phonon coupling is large enough \cite{EH76},
and they are always sharply localized as small bipolarons when
the interactions are local. Thus, taking physically
realistic parameters for the model, the effective mass of
the bipolarons becomes so huge (quasi-infinite) that it seems
quite unreasonable to expect the bipolarons to become
superfluid at a non-negligible temperature.
This aspect of the problem has been emphasized recently in
ref.\cite{CRF98}. However, the argument used by these
authors was based on standard considerations that did not 
take into account the effect of mass reduction we shall discuss in 
this and a subsequent paper \cite{PA99}.

Indeed, in realistic physical models, the characteristic energy of the bare
electrons is usually a few eV and is much larger than
the phonon energies which is at most about a tenth of an eV. 
As a result, the quantum fluctuations
of the phonons become generally negligible as soon as the electron-phonon
coupling is strong enough to generate bipolarons. Then
the potential interactions between the bipolarons are much larger
than their quantum kinetic energy. In that situation, the many bipolaron structures
should be well described by an effective Ising pseudospin Hamiltonian,
predicting an insulating Bipolaron Charge Density Wave 
at low temperature \cite{AQ89,AAR92,Aub93}.

However, there might exist special and exceptional situations
where the effective mass of the bipolarons is not quasi-infinite
but becomes small enough so that they  possibly condense into a superfluid state. 
The smaller the bipolaron mass is, the higher the critical temperature
should  be. As conjectured in ref.\cite{Aub93b} and \cite{Aub95b},
this situation might be produced by a well-balanced interplay between
the bare electronic kinetic energy, the electron-phonon coupling and the
direct electron-electron repulsion.  The aim of this paper is to study this
interplay in the simplest Holstein-Hubbard (HH) model where these interactions are present.

This first paper is devoted to the study of a single  bipolaron in
the HH model in the adiabatic limit, assuming classical phonons. 
Obviously the assumption
that there are no quantum phonon fluctuations does not allow
superfluid states (with many electrons).
In the next paper \cite{PA99}, the quantum phonon
correction to the adiabatic case will be studied. There, it will be
shown that in some regions of the parameter space,
there is indeed a drastic reduction of the quantum bipolaron's effective mass
due to quantum resonances between several
almost degenerate adiabatic bipolaron structures. A large part of
the scientific material of these two papers can be already found
(in French) in the PhD dissertation of one of us \cite{Pro98}.

Some numerical studies of the bipolarons in the one-dimensional 
adiabatic HH model, were already presented in ref.\cite{PA98}
(as well as few preliminary studies in two dimensions).
Bipolarons always exist in one-dimensional models as expected,
but when the Hubbard term $\upsilon$ increases from zero, a
first order transition occurs between the single site bipolaron (S0) and
a bipolaron (S1) composed of two bounded polarons on two neighboring
sites in a magnetic singlet state. It was observed that the
classical mobility of the bipolaron (assuming the
lattice dynamics is classical) was significantly enhanced
in the vicinity of this transition. 
Owing to the presence of the Hubbard term, quite small bipolarons
could become nevertheless highly mobile over hundreds of lattice spacings.

The behavior of the bipolaron in the two-dimensional case is quite different 
from the one-dimensional case. Although it does not describe precisely the
$CuO_2$ planes of cuprates \cite{Wal96}, it might exhibit similar features as
more realistic models. In two-dimensional models with local interactions, the
bipolarons exist only for a large enough electron-phonon coupling and are
always sharply localized (small bipolarons).  We numerically
calculate these bipolarons by using a continuation method of
these solutions from the anti-integrable limit \cite{Aub95}, where 
the electronic transfer integral is zero.

The ground state of the bipolarons in this limit can be easily found
and consists of either a bipolaron localized at a single site (S0) or
of two uncoupled polarons at arbitrary different sites, but there are
many other states with larger energy that are combinations of singlet states
(multisinglets). Many of these bipolaron states can be continued when the transfer
integral varies from zero and their energies can be compared.
Although the bipolaron (S0) or the singlet bipolaron (S1),  
persist with the lowest energy in large parts of the phase diagram, it is found that
a quadrisinglet state (QS) becomes the ground-state in an
intermediate regime of parameters.   

We show that we can reproduce quite accurately the same
phase diagram by choosing variational wave functions for the electrons
made from simple combinations of exponentials reproducing the main
characteristic of the spin structure of the bipolaron.
( This is an extension of the variational method used in  ref.\cite{KAT98}).
Further extensions could be developed later for the many-body problem.

We investigate the properties of all the obtained solutions by calculating
their binding energies, their pinning and breathing modes and  also their Peierls-Nabarro
energy barrier. We find a 
substantial softening of their pinning (and breathing) modes and a 
sharp depression of the PN energy barrier in the region where the (QS) bipolaron becomes
the ground-state. Although  the classical mobility 
of the bipolarons never becomes as large
as in the one-dimensional case \cite{PA98}, it is sufficient to favor a good quantum
mobility \cite{PA99} in a specific region of the phase diagram.

\section{The Model}

To keep in mind the physical magnitude of the dimensionless parameters
involved in our reduced model, let us first write the Holstein-Hubbard Hamiltonian
with all its parameters measured in the original physical units:

\begin{eqnarray}
\mathcal{H} = &-&T\sum_{<i,j>,\sigma}C_{i,\sigma}^{+}C_{j,\sigma}
+\sum_{i} \hbar \omega_0(a_{i}^{+}a_{i}) \nonumber \\
&& + \sum_i g n_{i}(a^{+}_{i}+a_i) + \sum_i
{\upsilon} n_{i,\uparrow}n_{i,\downarrow}   \label{hamiltonian}
\end{eqnarray}

The electrons are represented by the standard fermion 
operators $C_{i,\sigma}^{+}$  and $C_{j,\sigma}$ at site $i$ with 
spin $\sigma = \uparrow$ or $\downarrow$.
Then $T$ is the transfer integral of 
the electrons between nearest neighbor sites $<i,j>$ 
of the lattice. In physical systems, its order of
magnitude is usually measured in eV.

$a^{+}_i$ and $a_i$ are standard creation and annihilation boson 
operators of phonons. $\hbar \omega_0$ is the phonon energy of
a dispersionless optical phonon branch with order of magnitude
a tenth of an eV at most.

$g$ is the constant of the on-site electron-phonon coupling
which may physically range from zero to a fraction of an eV.
The on-site electron-electron interaction is represented by a
Hubbard term with positive coupling ${\upsilon}$ which may range
physically from negligible to large values of the order of 10 eV. 

Choosing $E_0=8 g^2/\hbar\omega_0$ as the energy unit
and introducing the position and momentum operators:

\begin{eqnarray}
&u_i&=\frac{\hbar\omega_0}{4g}(a_{i}^{+}+a_{i}) \label{position}\\
&p_{i}&=i\frac{2g}{\hbar\omega_0}(a_{i}^{+}-a_{i}) \label{momentum}
\end{eqnarray}

we obtain the dimensionless Hamiltonian:

\begin{eqnarray}
H = &&\sum_i{\left(\frac{1}{2} u_i^2 + \frac{1}{2}u_i n_i
+ U n_{i\uparrow}n_{i\downarrow} \right)}
-\frac{t}{2}\sum_{<i,j>,\sigma} C_{i,\sigma}^{+}C_{j,\sigma} 
\nonumber \\
&&- \frac{\gamma}{2} \sum_{i}p_i^2 \label{hamiltreduc}
\end{eqnarray}

The parameters of the system are now:
\begin{equation}
	E_0=8 g^2/\hbar\omega_0 \qquad  
	U = \frac{\upsilon}{E_{0}} \qquad 
	t =  \frac{T}{E_{0}} \qquad 
	\gamma = \frac{1}{4} (\frac{\hbar\omega_0}{2g})^4  
	\label{param} 
\end{equation}

The parameter $\gamma$ measures how "quantum" is the lattice. The BCS
theory requires $g << \hbar \omega_0$: that is, large $\gamma$.
We are interested in the opposite regime of strong electron-phonon
coupling: that is, $g$ larger than the phonon energy $\hbar\omega_0$.
Then $\gamma$ becomes small.
  
Thus the adiabatic approximation, which is simply obtained by taking $\gamma=0$,
becomes valid in the strong electron phonon regime. We shall assume this condition
in this first paper. Then $\{u_i\}$ commutes with the Hamiltonian and can be taken
as a scalar variable. For a given set of $\{u_i\}$, the adiabatic Hamiltonian 

\begin{equation}
H_{ad} = \sum_i{\left(\frac{1}{2} u_i^2 + \frac{1}{2}u_i n_i+
U n_{i\uparrow} n_{i\downarrow}\right)}
-\frac{t}{2}\sum_{<i,j>,\sigma}C_{i,\sigma}^{+}C_{j,\sigma} 
\label{hamiltelectr}
\end{equation}
commutes with the total spin of the system. 

Thus, the eigenstates of a system with two electrons are either nondegenerate
singlet states or three-fold degenerate triplet states. The wavefunction of the
singlet state has the form
\begin{equation}
|\Psi>=\sum_{i,j}\psi_{i,j}C_{i,\uparrow}^{+}C_{j,\downarrow}^{+}|
\emptyset>
\label{Psi_el}
\end{equation}
where $|\emptyset>$ is the vacuum (no electrons in the system) and
$\psi_{i,j}=\psi_{j,i}$ is normalized 

\begin{equation}
	\sum_{i,j} |\psi_{i,j}|^2 =1
	\label{normcd}
\end{equation}

on the $2D$ lattice $(\mathbf{Z}^D)^2$ ($D=2$ being the lattice
dimension we consider in this paper). The wave function of the triplet
state (oriented with the spin $+1$ in order to fix the ideas), has the form 

\begin{equation} 
|\Psi>_T=\sum_{i,j}\psi_{i,j}^T C_{i,\uparrow}^{+}C_{j,\uparrow}^{+}|
\emptyset>
\label{Psi_elT}
\end{equation} 

where $\psi_{i,j}^T=-\psi_{j,i}^T$ is normalized and antisymmetric.
Actually, the singlet wave and the triplet functions which are
eigenstates of the adiabatic Hamiltonian (\ref{hamiltelectr})  both yield
the same eigen-equation for their components $\psi_{i,j}$ or $\psi_{i,j}^T$
\begin{equation}
- \frac{t}{2} \Delta \psi_{i,j}  +
\left(\frac{1}{2} (u_i+u_j) + U \delta_{i,j}\right) \psi_{i,j} =
F_{el}(\{u_i\}) \psi_{i,j} 
\label{eleigen}
\end{equation}
where  $\Delta$ is the discrete Laplacian operator in the $2D$
lattice $(\mathbf{Z}^D)^2$ defined as
$ (\Delta \Psi)_i=\sum_{j:i} \Psi_j$ where $j \in (\mathbf{Z^D})^2$
are the nearest neighbors of $i \in (\mathbf{Z}^D)^2$.

Unlike the singlet states, the eigenenergies of the triplet states
do not depend on the Hubbard term $U$  since $\psi_{i,i}^T=0$
and thus are just the same as for noninteracting electrons.
Taking into account that in our model, the transfer integrals
with amplitude $t>0$ connect only the nearest neighbor sites,
it is straightforward to check that the singlet state defined as
$\psi_{i,j}=|\psi_{i,j}^T|$ always has less energy than the triplet state
with wave function $\{\psi_{i,j}^T\}$. As a result, the ground-state of our
system is necessarily a singlet state with the form (\ref{Psi_el}).

The energy of (\ref{hamiltelectr}) depends on $\{\psi_{i,j}\}$
and $\{u_i\}$ as

\begin{eqnarray}
F(\{\psi_{i,j}\},\{u_i\}) =\sum_i \left(\frac{1}{2} u_i^2
+\frac{u_i}{2} \rho_i + U |\psi_{i,i}|^2\right)&& \nonumber \\
-\frac{t}{2} <\psi|\Delta|\psi> &&  \label{eq3}
\end{eqnarray}

where the electronic density at site $i$ is 
\begin{equation}
\rho_i=\sum_j (|\psi_{i,j}|^2+|\psi_{j,i}|^2)
\label{rho_i}
\end{equation}

Extremalizing $F(\{\psi_{i,j}\},\{u_i\})$ with respect to the normalized
electronic state $\{\psi_{i,j}\}$ and the displacements $\{u_i\}$ yields
the set of coupled equations (\ref{eleigen}) and 

 \begin{equation}
u_i+\frac{\rho_i}{2}= 0   \label{NLEQ}
\end{equation}

$F_{el}(\{u_i\})$ is an eigenenergy of two interacting electrons in the
potential generated by the lattice distortion $\{u_i\}$. Using eq.\ref{NLEQ},
the extrema of eq.\ref{eq3} are those of the variational energy

\begin{equation} \label{Varform}
F_v(\{\psi_{i,j}\}) =\sum_i{\left(-\frac{1}{8} \rho_i^2
+ U |\psi_{i,i}|^2\right)} - \frac{t}{2} <\psi|\Delta|\psi>
\end{equation}
for $\psi_{i,j}$ normalized and where $\rho_i$ is given by eq.\ref{rho_i}.
Then, it follows that

\begin{equation}
- \frac{t}{2} \Delta \psi_{i,j} +
\left(-\frac{1}{4} (\rho_i+\rho_j) + U \delta_{i,j}\right) \psi_{i,j} 
= F_{el} \psi_{i,j} \label{Sceq}
\end{equation}
and also that the solutions of this equation, the energy of the system is

\begin{equation}
F_v(\{\psi_{i,j}\})=F_{el}+\frac{1}{8}\sum_i \rho_i^2
\label{Varform2}
\end{equation}

\section{Numerical Continuations of Bipolarons from the Anti-Integrable Limit}

\subsection{Bipolarons in the Anti-Integrable Limit}
\begin{figure*}
\begin{center}
\includegraphics [width=0.29 \textwidth]{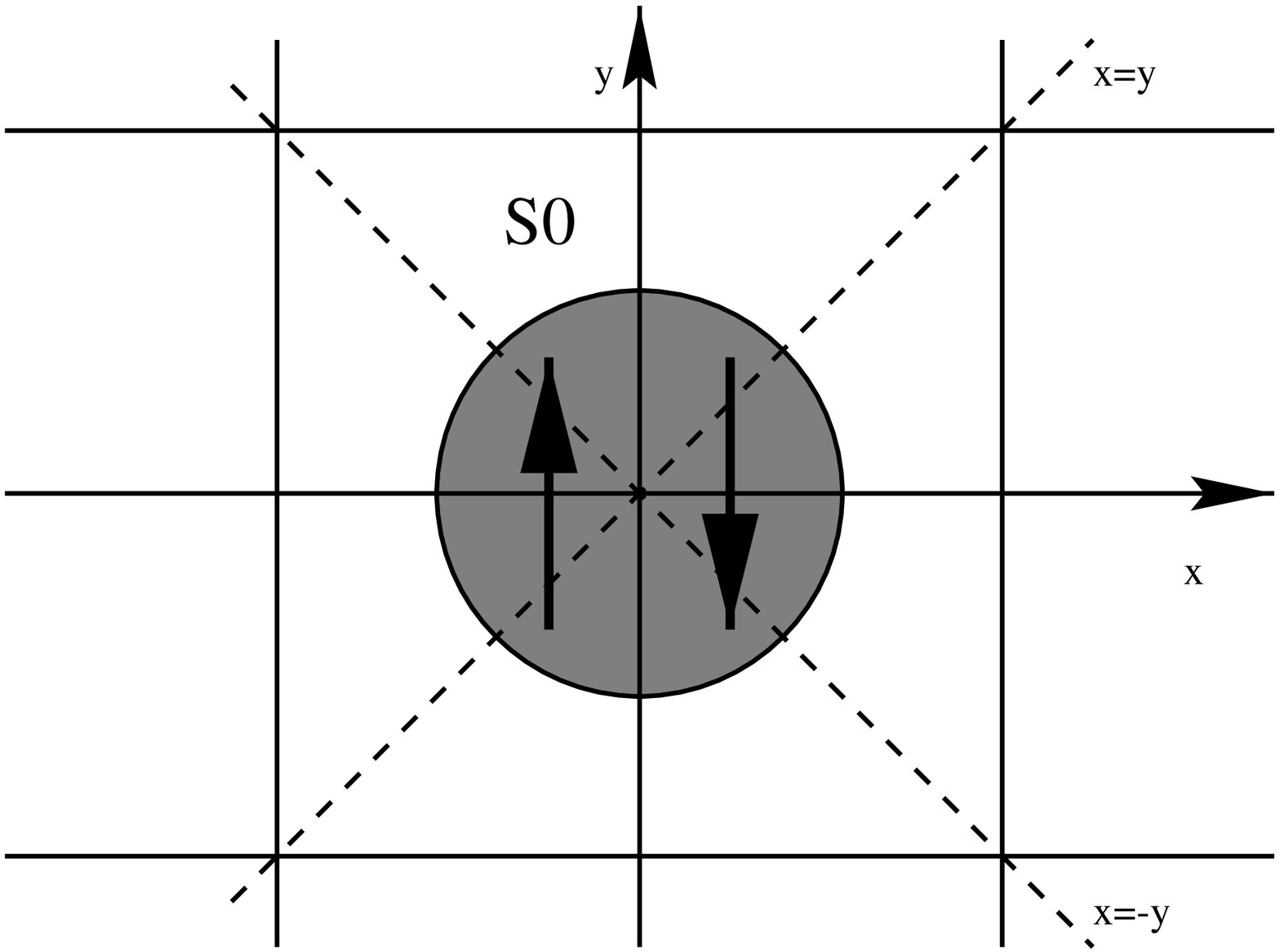}
\includegraphics [width=0.3 \textwidth]{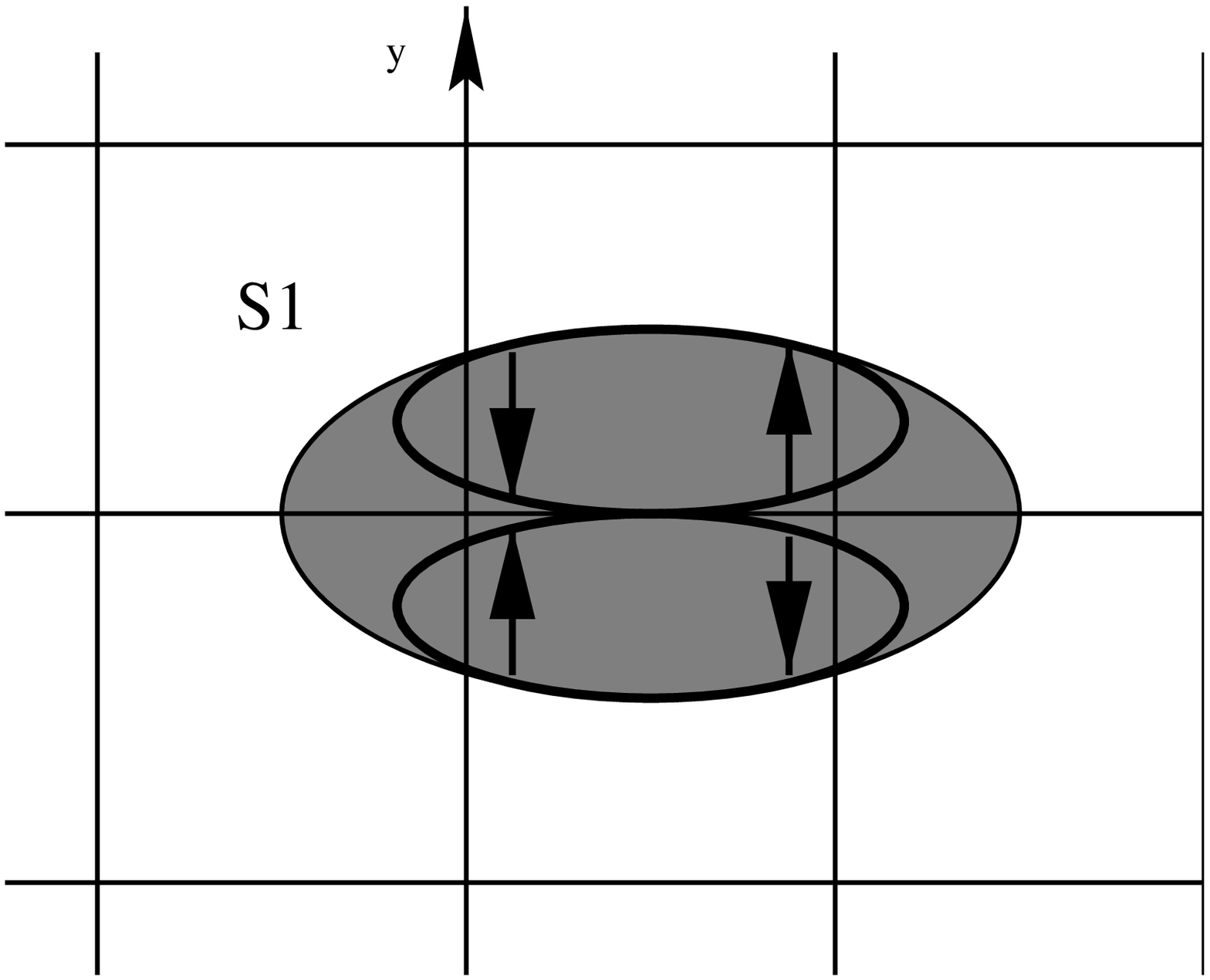}
\includegraphics [width=0.3  \textwidth]{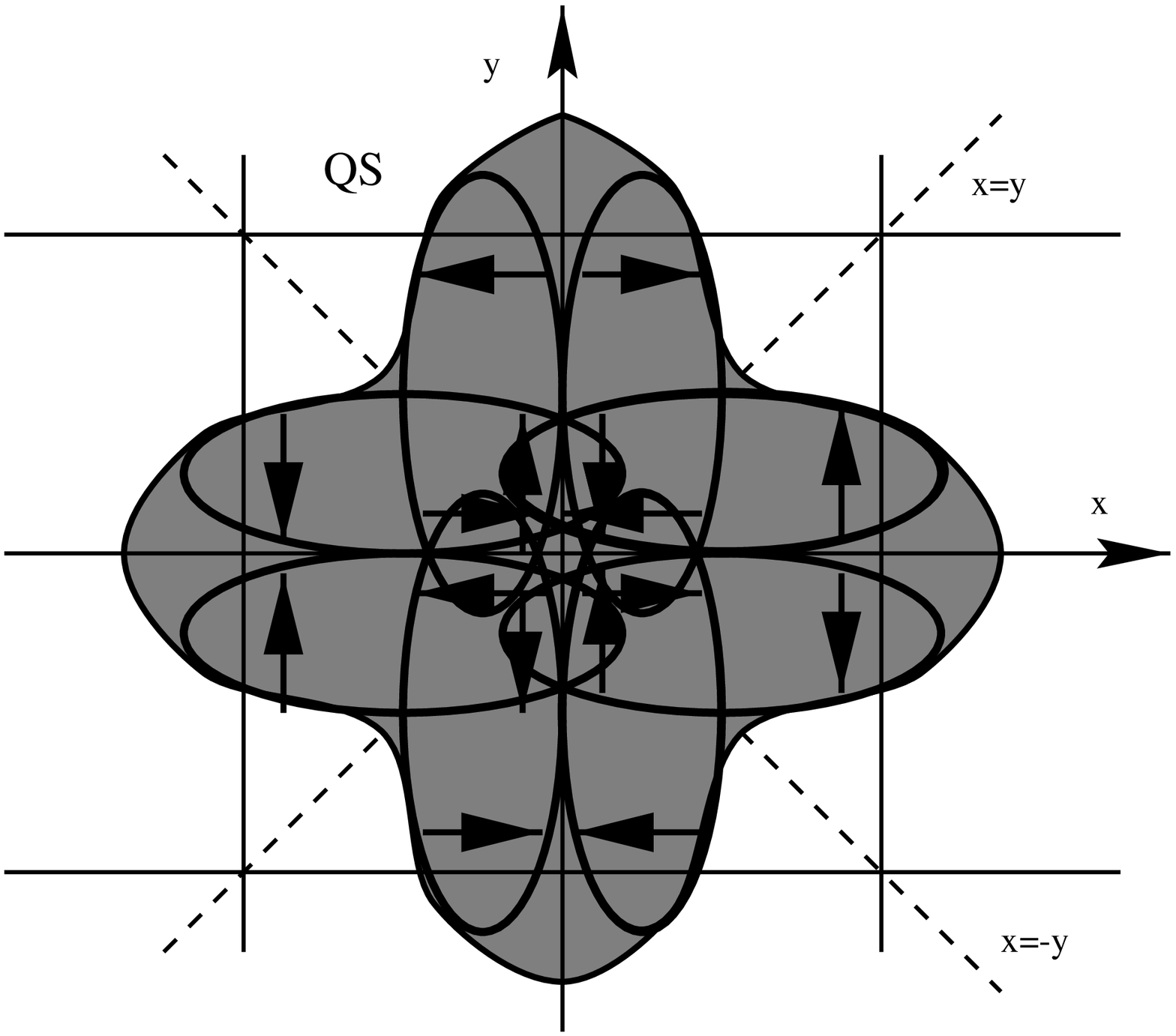}
\caption{ \label{fig:1}  
Schemes of the bipolarons (S0), (S1) and  (QS)
appearing as possible ground-states}\end{center}
\end{figure*}

In the anti-integrable limit $t=0$, the adiabatic ground-state for two electrons
is easily found. For $U \leq 1/4$, it consists of a pair of electrons
localized at a single site $i$. This is the standard small bipolaron 
known in the literature, denoted (S0) (see fig.\ref{fig:1}). For a
bipolaron at site $i$, its electronic wave function is

\begin{equation}
|\Psi> =  C_{i,\uparrow}^+ C_{i,\downarrow}^+ |\emptyset >
\label{bip0}
\end{equation}
and its energy is $F_v=U-1/2$.

When $U \geq 1/4$, the ground-state consists of two unbound polarons localized
at arbitrary different sites $i$ and $j$ and with arbitrary 
spins. It is thus degenerate and
its energy $F_v=-1/4$ is independent of the Hubbard interaction. 
When sites $i$ and $j$ are
nearest neighbors, we define the bipolaron (S1) \cite{Aub93b,Aub95b}
(see fig.\ref{fig:1}) with electronic wave function
 
\begin{equation}
|\Psi> = \frac{1}{\sqrt{2}}( C_{i,\uparrow}^+ C_{j,\downarrow}^+
+ C_{j,\uparrow}^+ C_{i,\downarrow}^+)|\emptyset > 
\label{bip1}
\end{equation}
where $i$ and $j$ are nearest neighbor sites.

Since a single polaron has the electronic spin $1/2$, when the transfer
integral $t$ is small but not zero, a standard perturbation theory yields
an antiferromagnetic exchange coupling $2t^2/U$  between the two spins of
the uncoupled neighboring polarons. 
When the spins are chosen in the singlet state represented by
eq.\ref{bip1}, these two polarons have the energy $F_v \approx -1/4 -t^2/U$.
When they are not located at nearest-neighbor sites but at the lattice distance
$n$,  perturbation theory to order $n$ yields an antiferromagnetic exchange coupling
proportional to $U(t/U)^{2n}$. Thus, for $t<<U$, the minimum energy is obtained for
nearest neighbor bipolarons in the singlet magnetic state (S1). It is maximum
when $U$ is close to and above $1/4$, just when (S1) becomes of lower energy than (S0).
For $t$ fixed, it decreases to zero when $U$ increases.
This binding energy also vanishes in the anti-integrable limit $t$.
Unlike bipolaron (S0), bipolaron (S1) breaks the
square lattice symmetry and is oriented either in the $x$ direction or the $y$ direction.

When $t$ is not very small, the spatial extension of the polarons
goes significantly beyond single sites, and it is not obvious that a low-order
perturbation theory holds. The true ground state might not be obtained by
continuation of the solutions (S0) or (S1). There are infinitely many other
bipolaron states at $t=0$ (solutions of eq.\ref{eleigen}), which have been
classified in appendix (\ref{AppA})
\footnote{Actually, this result should not be surprising since there are already
infinitely many metastable states in addition to the standard single bipolaron \cite{LeD83}
(see also section 5.4 in ref \cite{AAR92}) in the pure Holstein model, 
which however never become ground-state.}. Some of them are not very different 
in energy and
could compete to become the true bipolaron ground-state when $t$ increases.
Therefore, it becomes useful to test the ground-state of the bipolaron
at $t \neq 0$ among the extrema of eq.\ref{Varform} that are obtained
by continuation from those calculated in the anti-integrable limit at $t=0$.

It is of course impossible to continue and to test numerically  the energy of
all the solutions of eqs.\ref{eleigen} at $t=0$. The study of appendix (\ref{AppA})
shows that the binding energy of the bipolarons is non-negligible only when the
total number $N_s$ of occupied sites is not too large. The non-connected bipolaron
states are discarded because at $t=0$ they always have more energy than their
connected component with the smallest energy, and at $t \neq 0$, their absence of
connectivity is not favorable for gaining energy from the electronic kinetic
energy term with amplitude $t$.

On the contrary, the star multisinglet bipolaron states with one central site with
electronic density $\rho_1=1$ ($N_1=1$) and $N_2 \leq 4$ nearest neighbor sites with
electronic density $\rho_2=1/N_2$ and energy $F_v=-(1+1/N_2)/8$ at $t=0$ 
appear much more favorable for reducing their energy  when $t$ increases. They are still 
spatially well-localized,
which allows an efficient energy gain from the electron pho\-non coupling and only a small energy loss
due to the Hubbard term (no doubly occupied sites). Moreover, the peripheral electron
can gain a substantial electronic kinetic energy by occupying $N_2$ sites when $N_2>1$.
In the limit of $N_2$ large, this energy gain can reach a maximum at $2t$.
These bipolaron states have no continuous degeneracy at $t=0$ and thus according
to the implicit function theorem, they can be continued for $t$ not too large.

At $t=0$, the electronic wave function of a star $N_2$-singlet bipolaron 
centered at the origin $0$ is:

\begin{equation}
|\Psi> = \sum_{\nu} \frac{1}{\sqrt{2N_2}} (C_{0,\uparrow}^+
C_{j_{\nu},\downarrow}^+ + C_{j_{\nu},\uparrow}^+  C_{0,\downarrow}^+) |\emptyset >
 \label{multsing}
\end{equation} 

where $j_{\nu}$ are neighboring sites to the origin. They could also be chosen farther
away,but when the bipolaron becomes too extended, its energy does not decrease
sufficiently to become the ground-state. We tested the most compact bipolarons which are
star bisinglet bipolaron states (BS) with $N_2=2$ and $j_{\nu}$ 
are the two neighboring sites to the origin in the direction $x$ (or $x$ and $y$), star 
trisinglet bipolaron states (TS) where $N_2$=3 
and $j_{\nu}$ are three of the neighboring sites of the origin, and the
square symmetric quadrisinglet (QS) ($N_2=4$) which 
involves the four neighboring sites of the origin.

For larger or infinite lattices, multisinglets with equal electronic densities at the occupied
sites might not be too high in energy and have been also tested. Although they are continuously
degenerate in the anti-integrable limit, their degeneracy is raised
when $t \neq 0$. We considered for example, the square symmetric quadrisinglet state (QS2) which
occupies the four corners $j_{\nu}$ of an elementary square of the lattice.
One of its degenerate wave functions is 
\begin{equation}
|\Psi> = \sum_{\nu\neq \nu^{\prime}} \frac{1}{\sqrt{8}} C_{j_{\nu},\uparrow}^+
C_{j_{\nu^{\prime}},\downarrow}^+ |\emptyset >
\label{ringsing}
\end{equation}
with energy $F_v=-1/8$

\subsection{Numerical Technique of Continuation}

The most efficient numerical techniques for the continuation of solutions of sets
of equations as a function of a parameter, are usually based on a Newton method.
For example, such  techniques were developed efficiently for calculating discrete breathers
\cite{MA96}. In our case to calculate accurately adiabatic bipolarons on a 2D system, a
reasonable size should be $10 \times 10$. Then calculating the $10^4$ components of
$\psi_{i,j}$ with a Newton method, requires to work with huge matrices 
containing $10^8$ coefficients: that is, to use a large-memory, fast computer.
Actually, smaller size conventional computers suffice if one uses appropriate techniques needing
a much smaller working space.  This technique does not
allow to continue all solutions but only those which are locally stable
(in particular, the bipolaron ground-state) and those that can be made 
stable by fixing some spatial symmetries of the bipolaron.

\begin{figure*}
\begin{center}
\includegraphics[width=0.6 \textwidth]{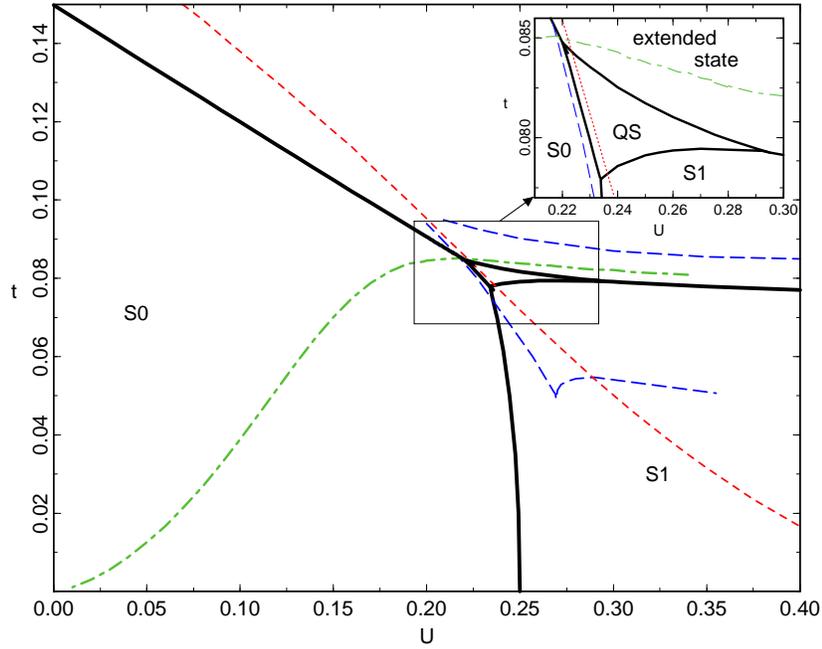}
\caption{\label{fig:2} 
Phase diagram of the bipolaron in the 2D 
Holstein-Hubbard model in the plane of parameters $U$ and $t$.  
There are four phase domains separated by first order 
transition lines corresponding to bipolarons (S0), (S1), (QS) and 
two unbound extended electrons. Also shown are the limit 
of metastability of the bipolaron (S0) (dotted line), 
 bipolaron (S1) (dot-dashed line) and bipolaron (QS) (dashed line).
Insert: Magnification of the phase diagram around the triple point
involving phases (S0), (S1) and (QS).}
\end{center}
\end{figure*}

This method is quite simple in its principle. To solve eq.\ref{Sceq} with condition
(\ref{rho_i}), we start from a normalized trial solution of eq.\ref{Sceq}, $\Phi=\{\phi_{i,j}\}$
with $\phi_{i,j}=\phi_{j,i}$, and we calculate recursively a new normalized 
trial solution
$\Psi_1= \mathcal{T}(\Phi) = \{\psi_{i,j}\}$ as

\begin{equation}
\mathcal{N}_1 \psi_{i,j} = - \frac{t}{2} \Delta \phi_{i,j} + \left(U \delta_{i,j} -
\sum_k (\phi_{i,k}^2+\phi_{j,k}^2) - K  \right) \phi_{i,j}
\label{algrth}
\end{equation} 
where $\mathcal{N}_1$ is the normalization factor (chosen negative) and $K$ is some positive constant
that we introduce to ensure the convergence to a minimum energy 
state. Actually, it can be chosen to be zero in the domain of parameter we study. 

We find numerically that for $n$ large enough, $\Psi_n= \mathcal{T}(\Psi_{n-1})$ 
and its normalization factor $\mathcal{N}_n$ converge to the limits $\Psi$ and
$\mathcal{N}$, respectively. $\Psi$ is a solution of eq.\ref{Sceq} with the condition
(\ref{rho_i}) and for the eigenenergy $F_{el}=\mathcal{N}-K$. This solution
corresponds to the  eigenvector of eq.\ref{Sceq} (where $\rho_i$ and $\rho_j$ are fixed)
associated with the eigenvalue $F_{el}$ which is such that  $F_{el}-K$ has the largest
modulus. In principle, the constant $K$ is chosen large enough in order that $F_{el}$ is surely
the lowest negative eigenvalue: that is, for the electronic ground-state.
One can easily check in the anti-integrable limit that $K=0$ is an appropriate
choice when $U<1/2$. Varying  one of the model parameters by small steps, each solution
is taken as a trial solution for the next step. It is easy to 
determine whether the
solution varies quasicontinuously or discontinuously.

For the solutions in the anti-integrable limit which are non-degenerate, it can be 
checked that the hypotheses of  the implicit function theorem, are fulfilled.
Thus continuation is in principle possible \footnote{The implicit function theorem
was already used in similar anti-integrable limits, for example in ref.\cite{BMK94}
for polarons and bipolarons in the original Holstein model or in ref.\cite{MA94}
for discrete breathers.}. For those which belong to a degenerate continuum, the
conditions for applying the implicit theorem are not fulfilled, but when
some spatial symmetries or some constraints on the solution are fixed, 
the degeneracy at $t=0$ can lifted and this theorem applies.

In the anti-integrable limit, only (S0) for $U<1/2$ and (S1) for $0<U$ (and (Sn)
with $n>0$ being the distance between two polarons) are numerically stable:
that is, can be followed continuously from $t=0$ by using algorithm (\ref{algrth}). Actually,
we choose as initial solution at $t=0$, the exact bipolaron solutions described
above, which are (S0), (S1), (QS), (BS), (TS) and (QS2). Maintaining by force
the spatial symmetries of the solution at $t=0$, the convergence process becomes stable
again, and the continuation of these solutions is feasible.

The main advantage of our method is that it can be performed on standard computers.
Its flaw is that we might not be able to follow continuously a solution 
that is mathematically continuable. Actually, rather few bipolaron states are continuable.
In contrast, our me\-thod is very reliable for finding the true bipolaron
ground state, because it brings spontaneously  the bipolaron solution to a local minimum
of the variational energy.

Actually, we checked that when there is no symmetry constrains and 
independant of the initial trial solution, in most cases our numerical algorithm converges
spontaneously toward the same bipolaron state, which then
can be considered as the true bipolaron ground-state.
However, this situation does not occur in the vicinity of the first order transition lines
where we can obtain a few  different bipolaron states depending on 
the initial condition,
but then their energies can be easily compared to find the ground-state.

\subsection{Bipolaron Phase Diagram}

The ground-state for a pair of electrons is obtained by comparing
the energies $F_v$ of many bipolarons continued from the 
anti-integrable limit (see also \cite{PA98}). For larger $t$, the ground-state corresponds to
a pair of electrons extended over the whole system. There is a first 
order transition line, when $t$ becomes smaller, at which the two 
electrons bind with each other and self-localize into a bipolaron. The region below 
this line is divided into three domains separated by other first-order transition lines.
For $U$ small, the bipolaronic ground-state is (S0). When $U$ 
increases for $t$ not too large, there is a transition line between bipolarons (S0) and (S1).
For larger $t$, this transition line bifurcates at a triple point at $t \approx 
0.785$ and  $U \approx 0.235$ into two first order transition lines 
which both join the transition line with the extended states.  
In between the fork that is generated, there is a small domain where the bipolaron (QS) 
that was initially unstable for $t$ small,  recovers its stability and even becomes
the ground-state. 
Other bipolaronic structures continued from the anti-integrable limit at 
$t=0$ appear as minimum energy states in the domain shown on 
fig.\ref{fig:2}. 
The (QS) solution can be viewed as a localized RVB state similar to that 
proposed by Anderson some years ago \cite{And87} in the pure Hubbard model
in 2D as a theory for superconductivity in cuprates. 

In our model, this (QS) bipolaron has the quantum symmetry $(s)$ 
because the kinetic energy term is Laplacian-like. 
However, the study of appendix (\ref{AppC}) in the anti-integrable limit, 
suggests that it is close in energy to other states with quantum 
symmetry $(s^{\prime})$ or $(d)$. Such symmetries could be favored 
by slight model variations on the form of the kinetic energy.

\begin{figure} 
\begin{center}
\includegraphics{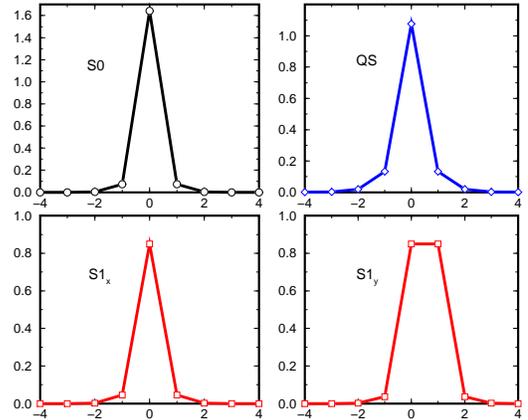}
\caption{\label{fig:3} 
Profile of electronic density versus site $i$ at the triple point $t=0.0779$, $U=0.234$
for bipolarons (S0), (QS) and (S1) along their symmetry $y$-axis and the 
transverse $x$-axis. These three bipolarons have the same energy.}
\end{center}\end{figure}

At the triple point, the bipolaronic structure of our model is degenerate between 
three states (S0), (S1) and (QS). Fig.\ref{fig:3} shows the profiles of the electronic density
for these three types of bipolaron, which have the same energy. Interestingly, 
they extend significantly over only a few sites, and thus can be called 
\textit{small bipolarons}.

\begin{figure}
\begin{center}
\includegraphics{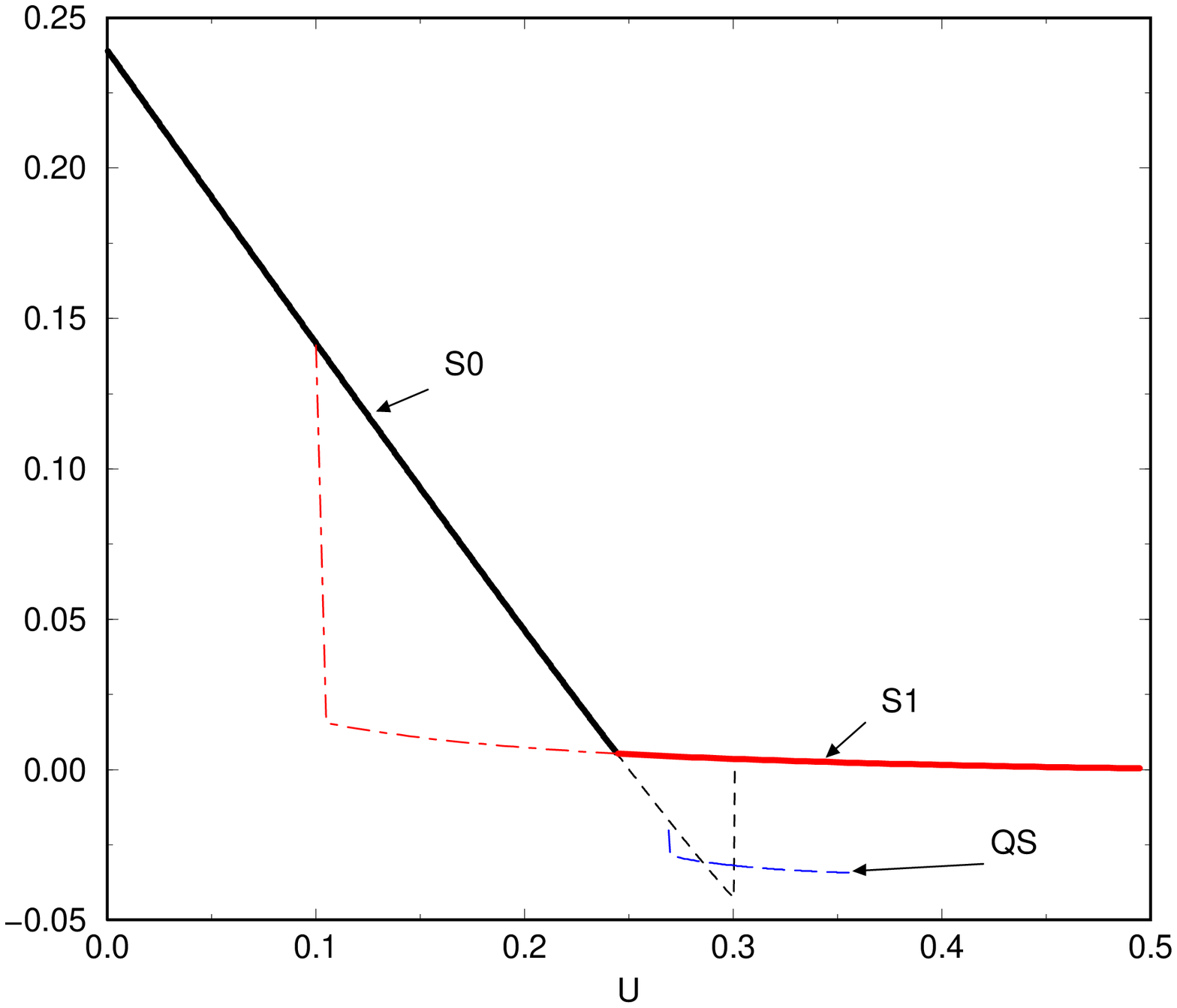}
\includegraphics{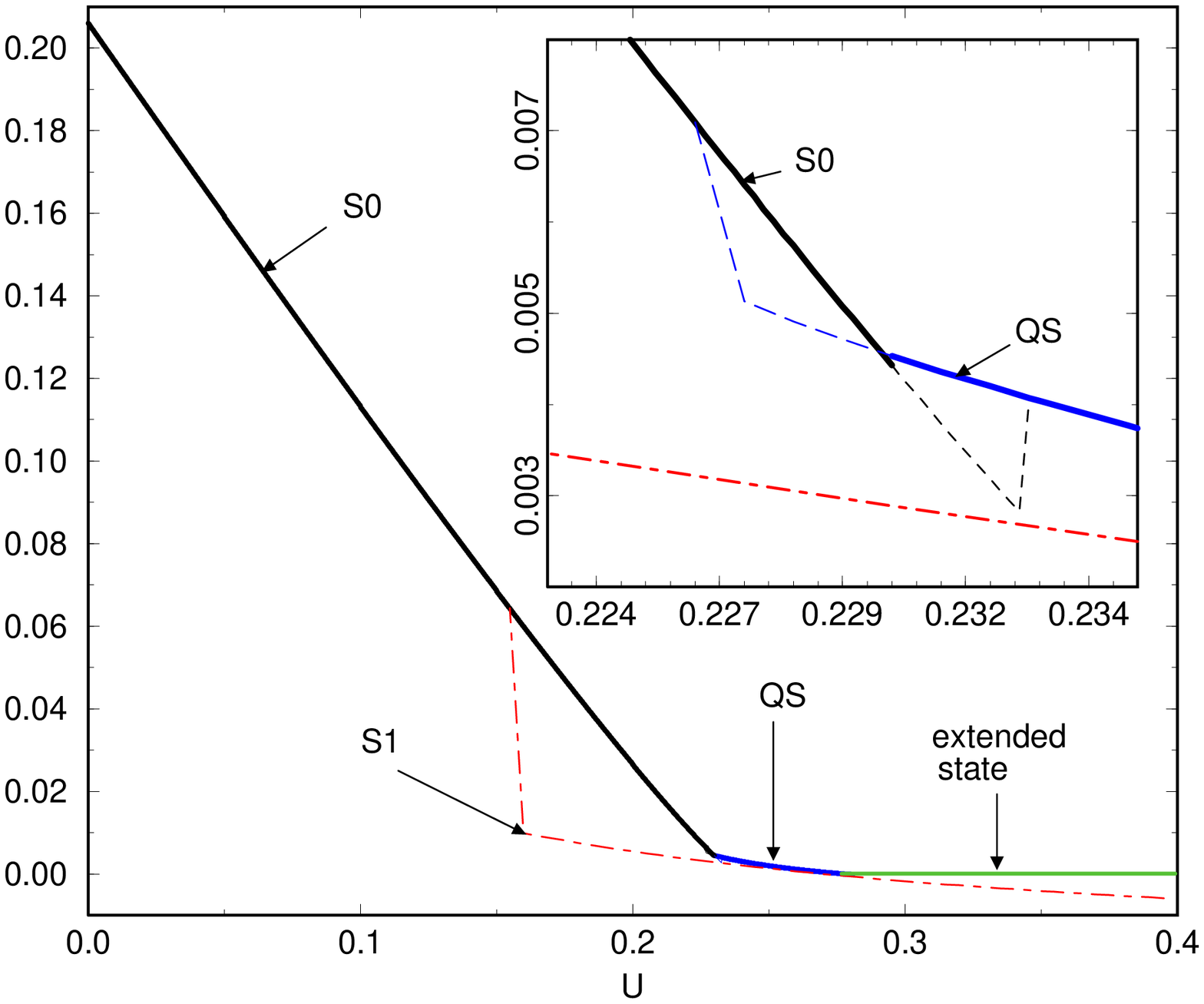}
\caption{\label{fig:4}  Binding energy versus $U$
of bipolaron (S0) (thin dotted line),
(S1) (thin dot-dashed line) and (QS) (thin dashed line) at $t=0.05$ (top)
(compared to two remote polarons) and $t=0.08$ (bottom)
compared to two extended polarons. The upper envelope (thick line)
is the binding energy of the ground-state.
Insert: magnification at the first order transition between  (S0) and (QS).}
\end{center}\end{figure}

The binding energy of a bipolaron is defined as the difference between the smallest energy 
state where the pair of electrons is unbound, and the bipolaron energy. Depending on the 
parameters, this unbound state could be either two extended electrons with 
opposite spins in the plane wave state at 
zero momentum or two polarons localized far apart.
The variation of this binding energy versus $U$ and for several 
values of $t$ is shown fig.\ref{fig:4}.

At the triple point,  the binding energy of the degenerate bipolarons 
(in that case, to produce two extended electrons) is much 
smaller than the binding energy of bipolaron (S0), at the same value 
of $t$ but at $U=0$. However, it still has a substantial value that is 
physically far from being negligible.

The binding of bipolarons (S1) and (QS) 
are physically better interpreted as being of magnetic origin.
These bipolarons can be viewed as two closely bound polarons with 
spins $1/2$. Their binding energy is mostly due to the spin energy gain 
obtained by lifting the spin degeneracy as a singlet state.

In the vicinity of the triple point and specifically in that region, the quantum lattice
fluctuations  ($\gamma \neq 0$) will also lift the degeneracy between the three degenerate bipolarons
(S0),(S1) (in both directions $x$ and $y$), (QS) 
resulting in a sharp mass reduction (or equivalently  a large tunneling 
energy or a large  band width) (see \cite{PA99}).

\section{Variational Calculation of Bipolarons}
\label{sec4}
We now reproduce, with good accuracy, the phase diagram shown in fig.\ref{fig:2}
using simple variational approximations for the bipolarons (S0), (S1) and (QS).
For that purpose, the variational forms  have to be chosen appropriately
under two conflicting constraints. On the one hand, 
they should be physically realistic enough in order
to mimic the real ground-state. On the other hand, the analytical calculations of their
variational energy should be practically feasible.  

In ref.\cite{KAT98}, it was shown that an exponential form centered at the occupied site 
with a unique variational parameter, was a good variational form for a single polaron,
reproducing accurately its quantitative properties. We choose a similar normalized variational
form for the electronic wave function of bipolaron (S0) located at the origin

\begin{equation}
\psi_{i,j}^{S0}=A \lambda^{(|i|+|j|)}\ 
\mbox{with} \ A=(\frac{1-\lambda^2}{1+\lambda^2})^2
\label{AS0}
\end{equation}
 (for $i=(i_{x},i_{y})$, we set $|i|=|i_{x}|+|i_{y}|$).
This variational form is easily extended to the electronic wave function of bipolaron (S1) 
in a singlet magnetic state located at sites $(0,0)$ and $(1,0)$:
\begin{eqnarray}
\psi_{i,j}^{S1}&=&\frac{B}{\sqrt{2}} (\lambda^{(|i_x-1|+|i_y|+|j_x|+|j_y|)}+
\lambda^{(|i_x|+|i_y|+|j_x-1|+|j_y|)}) \nonumber\\
\quad  &&  \mbox{with} \quad 
B=\frac{(1-\lambda^2)^2}{(1+\lambda^2)\sqrt{1+6\lambda^2+\lambda^4}}
\label{AS1}
\end{eqnarray}

The variational form for the electronic wave function of bipolaron (QS) centered at the origin
is a combination of four of these variational forms in the four directions of the square
lattice, but now it becomes useful to introduce two variational parameters  $\lambda$ and $\mu$
instead of only one, to distinguish between the spatial extension of the polaron
that is at the center from those that are the periphery:

\begin{eqnarray}
&&\psi_{i,j}^{QS}=\frac{C}{\sqrt{8}} 
 \mu^{(|j_x|+|j_y|)} \sum_\pm (\lambda^{(|i_x \pm 1|+|i_y|)}+\lambda^{(|i_x|+|i_y\pm 1|)})
\nonumber\\
&+ &\frac{C}{\sqrt{8}}  \mu^{(|i_x|+|i_y|)} \sum_\pm
(\lambda^{(|j_x\pm 1|+|j_y|)}+\lambda^{(|j_x|+|j_y\pm 1|)} ) 
\label{ASQ}
\end{eqnarray}

where for normalization
\begin{eqnarray}
C^{-2}&=& (\frac{1+\mu^2}{(1-\mu^2)(1-\lambda^2)})^2 \nonumber \\
& & \left[(1+\lambda^2)^2 
+\lambda^2(3-\lambda^2)(1+\lambda^2)+8\lambda^2\right] \nonumber\\
 & & +4\frac{(1+\lambda\mu)^2(\lambda+\mu)^2}{(1-\lambda\mu)^4} 
 \label{Cnorm}
\end{eqnarray}

\begin{figure} 
\begin{center}
\includegraphics{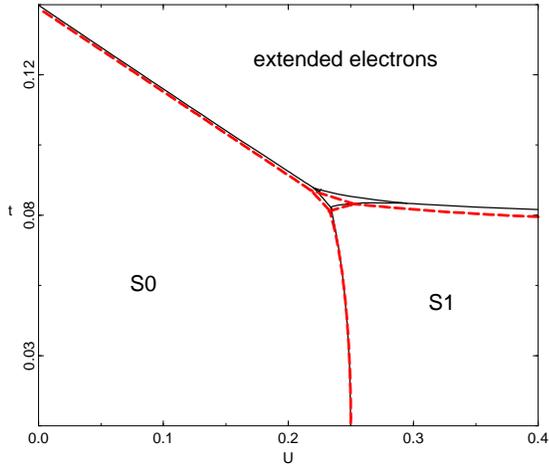}
\caption{\label{fig:5}   Comparison between the exact phase diagram 
fig.\ref{fig:2} (thick lines) and its approximate 
calculation (dashed lines)  sec.\ref{sec4}}
\end{center}\end{figure}

The energy (\ref{Varform}) can be analytically calculated with the variational
forms (\ref{AS0}), (\ref{AS1}) and (\ref{ASQ}). Extremalizing the 
resulting energy with respect to 
the parameters $\lambda$ and $\mu$ yields the energies of bipolarons 
(S0), (S1) and (QS) with a very good accuracy. We do not reproduce 
here these tedious calculation. We also remark that this variational method allows one
to compute the bipolaron structures
even when they become unstable so that they cannot be numerically 
continued with our method. 
Comparing these variational energies allows one to produce a phase 
diagram that is very close to the exactly calculated one (see 
fig.\ref{fig:5}).

However, it is worthwhile to mention that the variational form (\ref{ASQ}) of bipolaron
(QS) may yield some artefacts which are not found in the exact numerical calculations
(as often occurs in variational calculations). Fortunately, they occur in parameter
regions where this solution is not the ground-state, and thus do not  affect the phase diagram.
First, there is a first order transition of $\lambda$ and $\mu$
near the anti-integrable limit. Second, there is another anomaly when increasing $U$.
It is found that $\psi^{QS}$ bifurcates onto a unbound solution where 
$\mu=1$. This corresponds to an unbound  pair of electrons in the spin singlet state
where one electron is self-localized as a polaron and the second one is extended. 
As a result, the validity of the exponential form $\psi^{QS}$ is limited
to the (QS) region, that is when this bipolaron is the ground-state.

\section{Internal Modes and Peierls Nabarro Barriers}
\label{PNpart}

\begin{figure} \begin{center}
\includegraphics{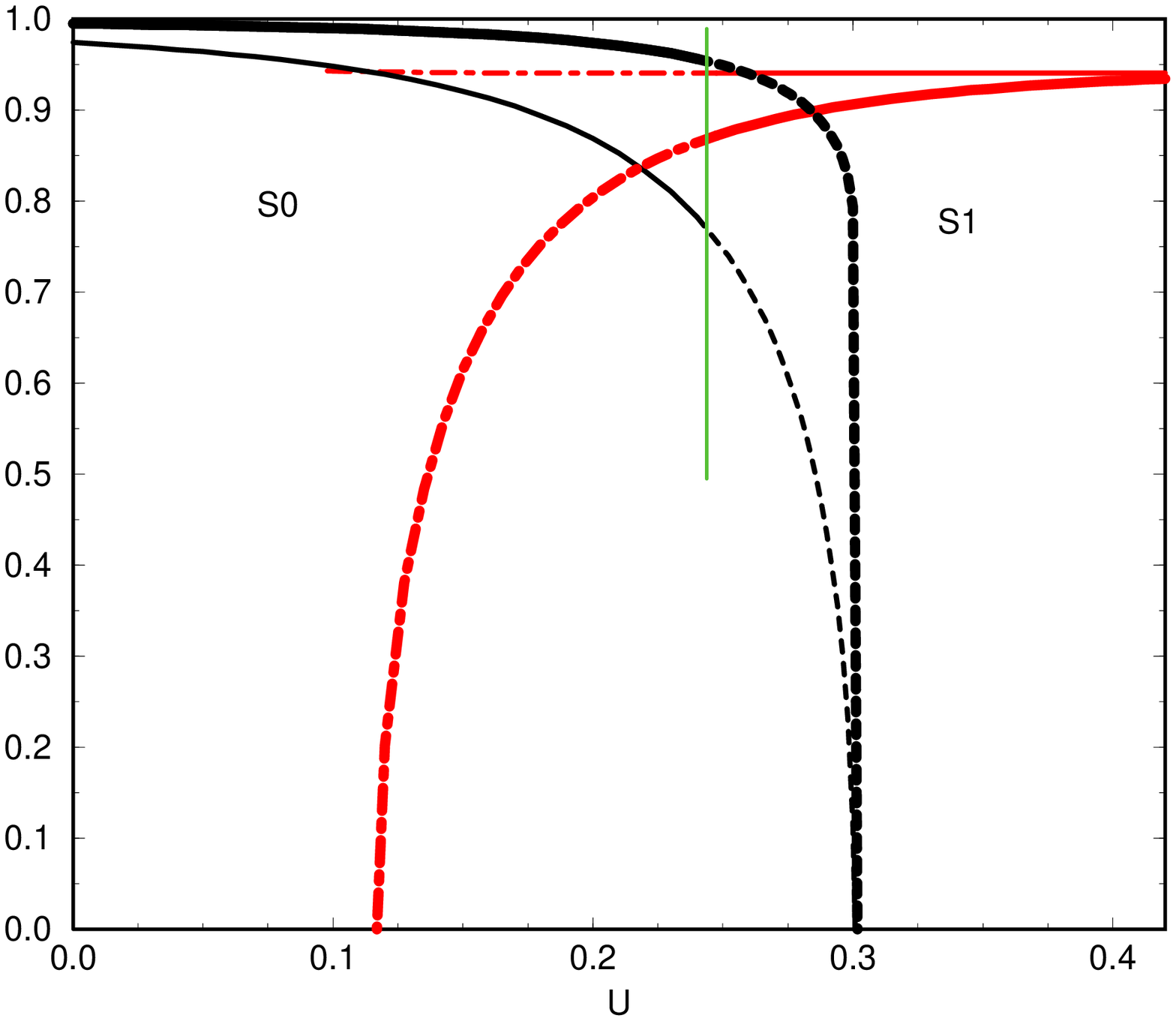}
\includegraphics{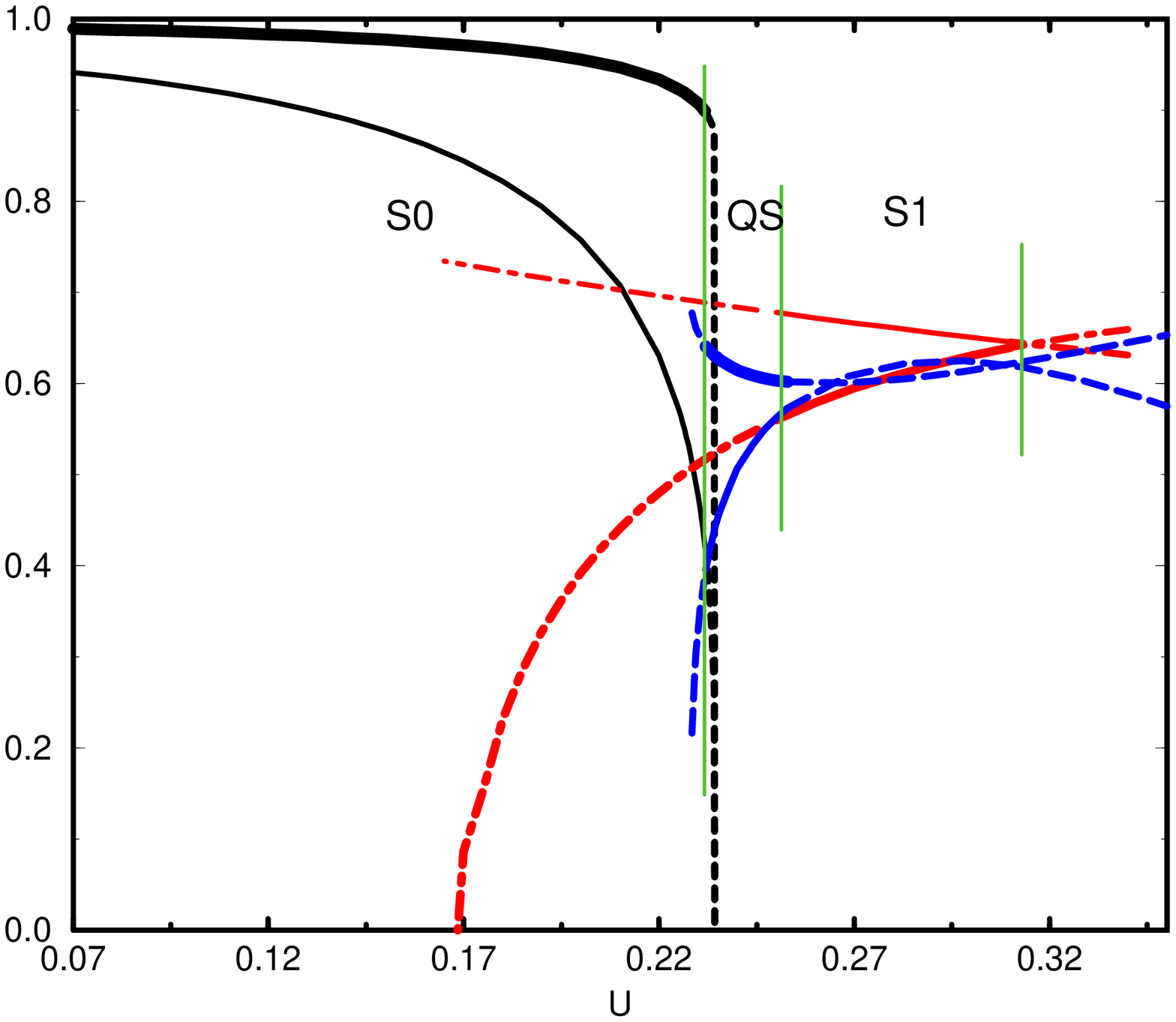}
\caption{\label{fig:6} 
Phonon frequencies versus $U$ of the pinning mode (thick line) and 
breathing mode (thin line) for bipolaron (S0) (dotted line), bipolaron (S1) (dot-dashed line), 
 bipolaron (QS) (dashed line) at $t=0.05$ (top) and $t=0.08$ (bottom). 
When, these bipolarons are ground-state, the corresponding lines are 
plain. Vertical lines indicate the first order transitions.}
\end{center}\end{figure}

The phonon frequencies of the bipolaron can be easily calculated within the standard
Born-Oppenheimer approximation (in units $\sqrt{\gamma}$)  as explained in \cite{PA98}. It is 
found that the bipolarons exhibit  several localized (or internal) modes.
The breathing mode has the same symmetry as the bipolaron. The pinning modes are spatially
antisymmetric and tend to move this bipolaron either in the $x$ direction or the $y$
direction. Fig.\ref{fig:6} shows the variations of their frequencies with $U$. It is found that
in the region of the triple point where three bipolaronic structures are almost degenerate,
both the breathing and the pinning modes, soften significantly (approximately by a factor $2$).

These weak frequencies for the internal modes can be considered as evidence 
that the self consistent potential in which the bipolaron is pinned 
becomes rather flat, which means a small Peierls Nabarro barrier (PN).
It is thus useful to calculate precisely  this PN energy barrier in 
order to confirm this conjecture. In addition, it is found that the 
paths that yield the lowest PN energy barrier vary in the parameter space.
The several ways to move the bipolaron are sometimes 
almost equivalent in energy. These paths should play a role in the quantum
tunnelling of the bipolarons

The PN energy barrier is the minimum energy that must be provided to the bipolaron to move
it by one lattice spacing. For that we have to determine a continuous path of bipolaronic
configurations which connects the initial bipolaron to a shifted equivalent bipolaron.
There is a maximum of energy along any path, and the minimum over all 
paths of this maximum (called minimax) yields the PN energy barrier.

To move a bipolaron with electronic wave function $\{\psi_{n,m}^{i}\}$ 
from site $i$ to a neighboring site $j$ (where $\{\psi_{n,m}^{j}\}= \{\psi_{n+j-i,m+j-i}^{i}\}$),
we consider a continuum of bipolaronic solutions  $\{\psi_{n,m}(c)\}$
which depend on $c$ for $c_{0}\leq c \leq c_{1}$, and such that
$\{\psi_{n,m}(c_{0})\}=\{\psi_{n,m}^{i}\}$ and 
$\{\psi_{n,m}(c_{1})\}=\{\psi_{n,m}^{j}\}$. 

It is convenient for simplicity to choose as variable $c(\Psi)$, one of the bipolaron 
components or a simple function of them. For any continuous path that connects the bipolaronic
ground-state $\Psi^{i}=\{\psi_{n,m}^{i}\}$ at site $i$ to the same configuration
$\Psi^{j}=\{\psi_{n,m}^{j}\}$
at an equivalent neighboring site $j$, $c$  must take all the values between $c_{0}=c(\Psi^{i})$
to $c_{1}=c(\Psi^{j})$. For each value of  $c$, the energy of the bipolaronic state will be
always larger than or equal to the minimum of energy of the bipolaronic configuration where the 
component corresponding to $c(\Psi)$ is fixed to $c$. Thus, starting from the initial ground-state
configuration, and following continuously this minimum by varying this constraint $c$,
we may pull continuously the bipolaron from one site $i$ to its neighboring site $j$.

For that purpose,  the choice of $c(\Psi)$ has to be appropriate to
obtain a path of bipolaronic configurations that connects \textit{continuously}
the two bipolaronic configurations $\Psi^{i}$ and $\Psi^{j}$ and that 
yields the lowest minimax. We guess intuitively that the bipolaron could be effectively
pulled only if this constraint affects the "main body" of the bipolaron instead of a minor component.
For our investigations, we found several continuous paths of configurations competing for providing
the minimax. We obtain them by using several kinds of constraints
for a bipolaron at site $i$ moving to site $j$ which may be:

\begin{eqnarray}
	\psi_{i,i} &=& c   \label{pthvar1}\\
	\psi_{i,j} &=& \psi_{j,i} =c \label{pthvar2} \\
	\psi_{j,k} &=& \psi_{k,j} = c \label{pthvar3} \\
	\psi_{i,i} &-& \psi_{j,j} = c \label{pthvar4}
\end{eqnarray}
($k \neq i$ is a neighboring site of $j$ and bond $j-k$ could be 
either collinear with or orthogonal to bond $i-j$.).
These constraints $c$ are easily taken into account with few 
minor changes in the numerical programs described above minimizing the variational form
(\ref{Varform}). When $\psi_{i,i}$, $\psi_{i,j}$ or $\psi_{j,k}$ is fixed to $c$, it suffices to drop
the corresponding equation (\ref{Sceq}). When $\psi_{i,i} - \psi_{j,j}= c$, it is convenient to 
define the new variable  $\phi=(\psi_{i,i}+\psi_{j,j})/\sqrt{2}$.
Then,  the variational form (\ref{Varform}), depends on $\phi$, $c$ and
$\psi_{n,m}$ for  $(m,n) \neq (i,i)$ and $\neq (j,j)$. 
The vector with components $\phi$ and $\{\psi_{m,n}\}$ (with $(m,n) \neq (i,i)$ 
and $\neq (j,j)$), has norm $\sqrt{1-c^{2}/2}$ which is fixed by 
the constraint $c$. Extremalizing (\ref{Varform}) with respect to its free variables
yields a set of equations that differ slightly from (\ref{Sceq}),
although they depend on $c$ as a parameter. They can be solved with the same
iterative method as before.

We may thus obtain a continuous path of configurations  parameterized by $c$ and connecting the
bipolaron ground-state to the same state shifted  by one lattice spacing.
The extrema of the energy $F_{v}(c)$ given by (\ref{Varform}) (which satisfy 
$\partial F_{v}(c)/\partial c = 0$) correspond to bipolaronic solutions
without constraint. Actually, they can be identified among the bipolarons
that are classified in appendix (\ref{AppA}).  When one of these  bipolaron is found spatially
symmetric with a symmetry center at the middle of the bond $<i,j>$, 
there is no need to continue the path beyond this point because it is clear 
that it can be completed by symmetry. We test the different constraints (\ref {pthvar4})
and  among those that yield a continuous path, the lowest maximum is
considered as the minimax.

The PN energy barrier is then the difference between the minimum of energy and the
maximum along this best continuous path.

When the PN energy barrier is smaller than or at most comparable to
the binding energy of the bipolaron, the two polarons remain surely bound during
their continuous lattice translation and it is then reasonable to 
believe that the minimax obtained with the above method is correct. 
However, there are regions in the parameter space where this condition is not fulfilled, and
a precise determination of this PN energy barrier might be questionable. In any case, 
the obtained value, if not exact, is necessarily an overestimate.

\begin{figure} \begin{center}
\includegraphics{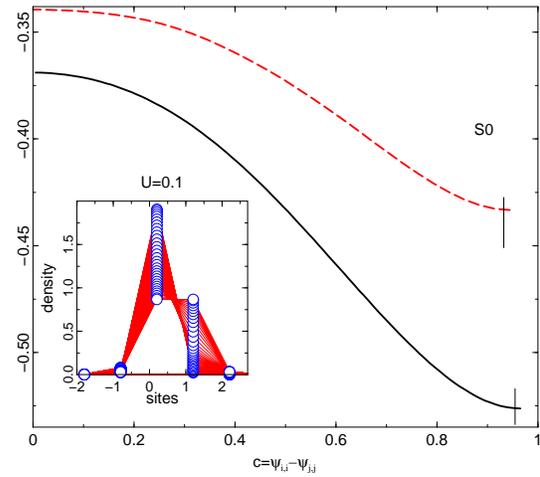}
\caption{\label{fig:7} 
Energy variation versus $c$ of the bipolaronic ground-state state under the 
constraint $\psi_{i,i}-\psi_{j,j}=c$. Bipolaron (S0) is initially at site 
$i$ and  $j$ is the neighboring site towards which this bipolaron moves. 
$t=0.08$ and $U=0$ (plain line), $U=0.1$ (long dashed line).
Vertical lines indicate the location of the energy extrema corresponding to
the initial bipolaron (S0). 
Insert: Variation of the Profile of electronic density along the 
continuous  path of the bipolaron at $t=0.08$, $U=0.1$.}
\end{center}\end{figure}

When (S0) is the ground-state and $U$ is not too large, the energy variation versus
$c=\psi_{i,i}-\psi_{j,j}$ starting from the ground-state bipolaron (S0) is shown in fig.\ref{fig:7}.
This path does not need to be continued beyond the point $c=0$ because it exhibits a minimax
at $c=0$ corresponding to a spatially symmetric bipolaron and thus can be completed by symmetry. 
This bipolaron has only one unstable mode and is the continuation at $U$ small
of the stable bipolaron (S1) beyond its bifurcation point.
At $U=0$, its electronic state corresponds to two electrons with 
opposite spins in the lowest eigenstate of the potential generated by 
the lattice distortion (Slater determinant). The analysis of appendix
(\ref{AppB}) suggests that for $U<0$, this bipolaron can be continued
as the two-site bipolaron (2S0) in the anti-integrable limit. 

\begin{figure} \begin{center}
\includegraphics{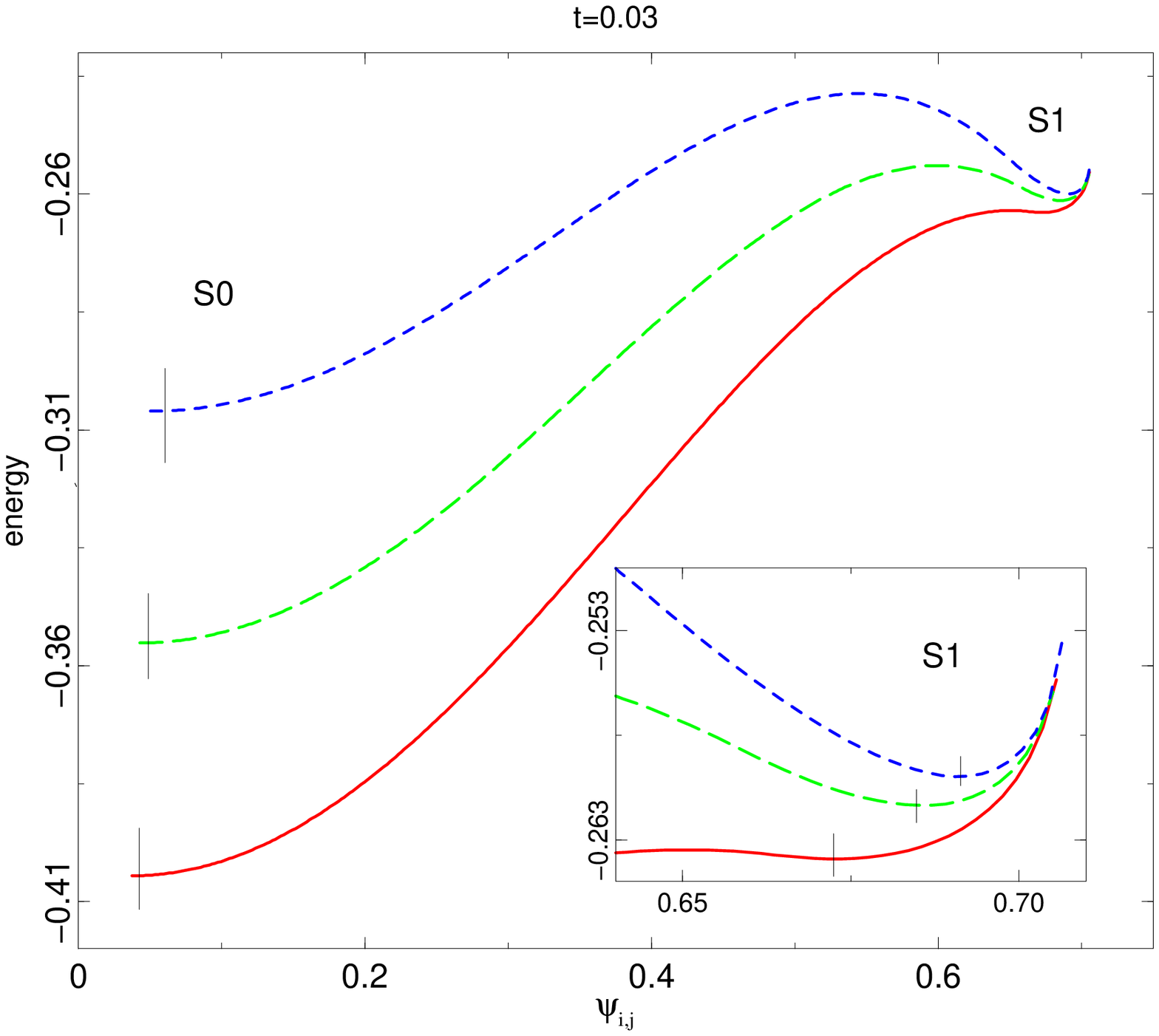}
\includegraphics{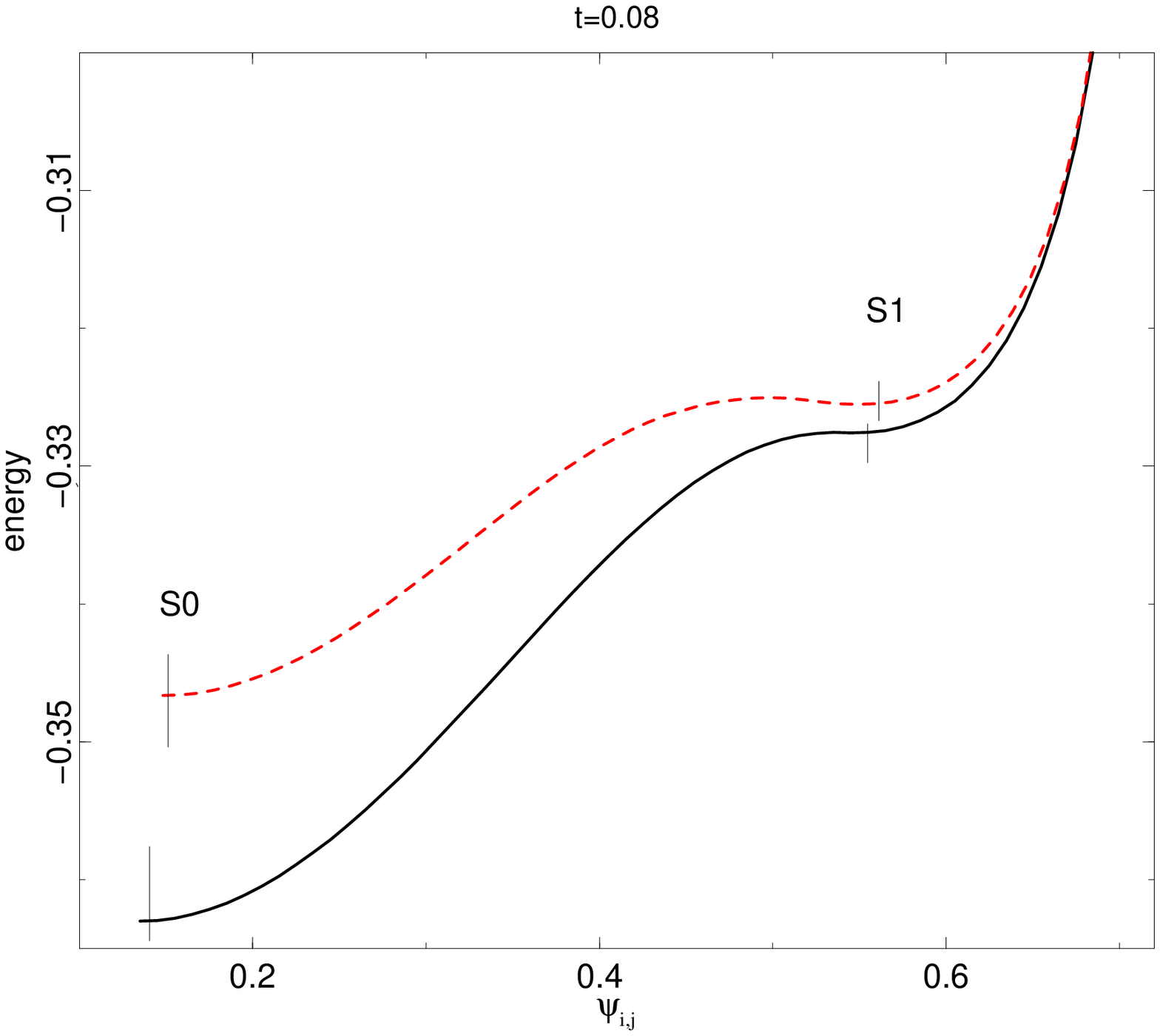}
\caption{\label{fig:8} 
Same as fig.\ref{fig:7} but with the constraint $\psi_{i,j}=c$
and different parameters.
top: $t=0.03$,  $U=0.1$ (full line),
$U=0.15$ (long dashed line), $U=0.2$ (dashed line)
insert: magnification
bottom: $t=0.08$, $U=0.18$ (full line), $U=0.2$ (dashed line).
Vertical lines indicates the location of the energy extrema corresponding to bipolaron 
(S0) and to bipolaron (S1).}
\end{center}\end{figure}

When $U$ becomes larger, the previous constraint does not always yield 
a continuous path, and another constraint $\psi_{i,j}=c$ is 
found efficient for providing a continuous path with a minimax. The energy variation versus
$c= \psi_{i,j}$ starting from the ground-state bipolaron (S0) is shown in fig.\ref{fig:8} for some 
bipolarons. We observed that in addition to the minimum (S0), it exhibits two 
other extrema.  The second minimum corresponds to the spatially symmetric bipolaron
(S1). Again, there is no need to construct a complete path reaching (S0), since this
path can be completed by symmetry. This figure shows that a pitchfork bifurcation
occurs for the minimax when $U$ increases from zero (at fixed $t$). The 
unstable bipolaron (2S0) bifurcates into a minimum corresponding to the stable bipolaron 
(S1) and two symmetric minimax corresponding to intermediate unstable bipolarons (with one unstable 
mode), which are nothing but the star sister bipolarons (S1/S0) described in appendix (\ref{AppA3})
\footnote{It is worthwhile to note a similar phenomenon observed when narrow
discrete breathers become mobile \cite{AC98}. Intermediate discrete 
breathers breaking the lattice symmetry were also found to appear
\cite{AC99}.}. Actually, this bifurcation line between (S1/S0) and (S1)
appears on fig.\ref{fig:2} as the left border line of the domain of metastability of 
bipolaron (S1).
In that regime, the motion of the bipolaron involving the minimum 
energy consists in first stretching bipolaron (S0)
into bipolaron (S1) along one lattice direction, and next in squeezing this bipolaron in the same 
direction to recover the bipolaron (S0) translated by one lattice 
spacing. This feature is identical to those found for the two-site model 
in appendix (\ref{AppB}).

\begin{figure} \begin{center}
\includegraphics{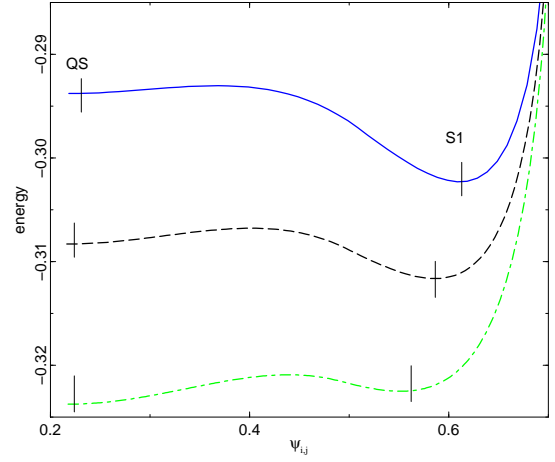}
\caption{\label{fig:9} Same as fig.\ref{fig:7} but for bipolaron (QS) 
with the constraint $\psi_{i,j}=c$ at 
 $t=0.08$, $U=0.235$ (dotted-dashed line),
$t=0.075$, $U=0.25$ (dashed line) and $t=0.07$, $U=0.26$ (full line),
(Vertical lines indicates the location of the energy extrema 
corresponding to bipolaron (QS) and (S1).}
\end{center}\end{figure}

When  bipolaron (QS) (which does not exist for the two-site model)
becomes the ground-state instead of (S0), the PN energy barrier should be studied from this
initial configuration. 
Fig.\ref{fig:9} shows the energy variation versus $\psi_{i,j}=c$
starting from bipolaron (QS).  The continuous path exhibits another
minimum corresponding to the stable bipolaron (S1), which is spatially 
symmetric. Again, the continuous path can be completed by symmetry.
There is a minimax which correspond to another bipolaronic configuration, which
we did not analyze in detail but is likely to be the star trisinglet 
denoted (TS) described in appendix (\ref{AppA2}). This curve also demonstrates that for
this value of $t$, the bipolaronic ground-state changes by a first-order transition from
(QS) to (S1) when increasing $U$.

\begin{figure} \begin{center}
\includegraphics{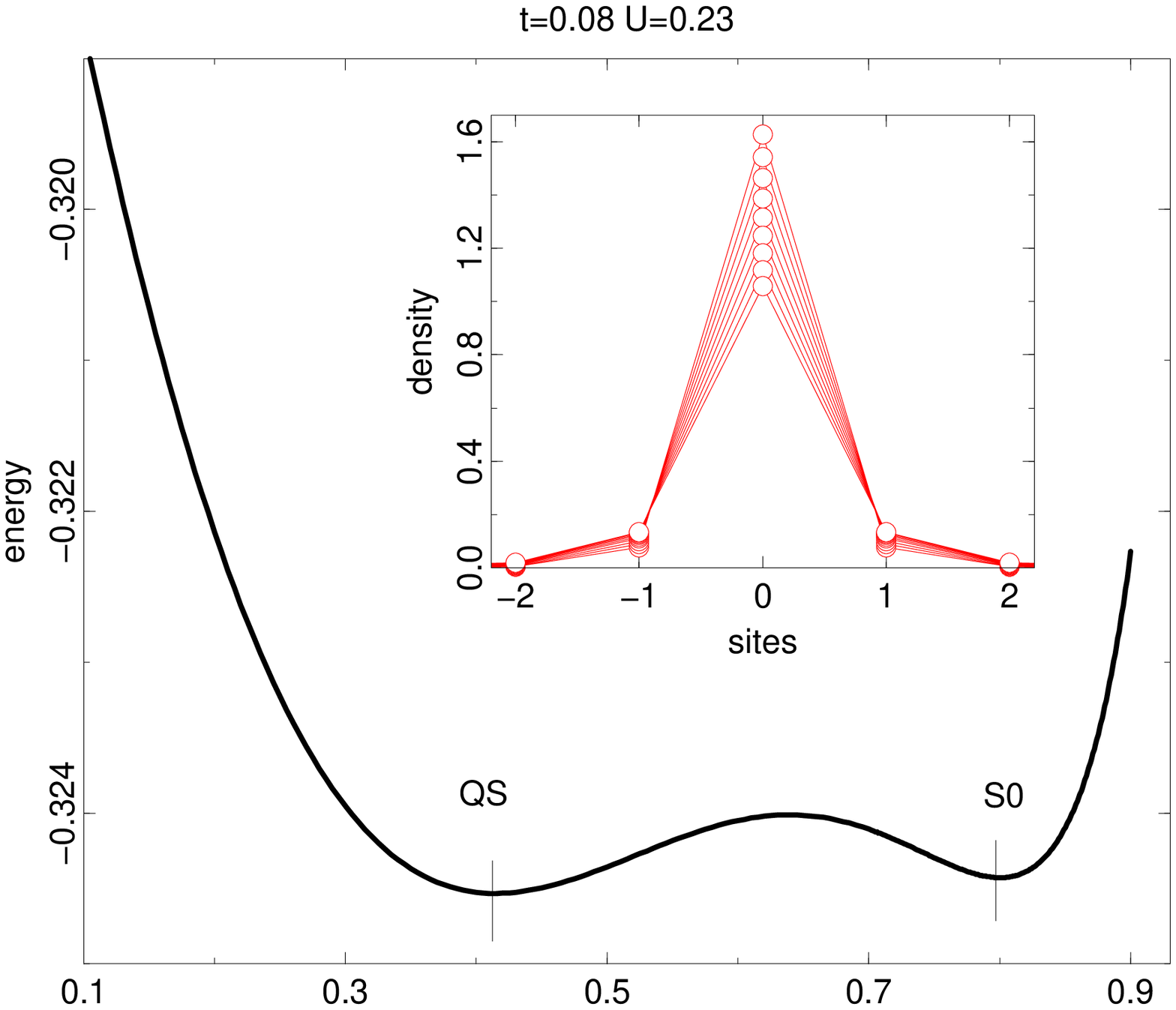}
\includegraphics{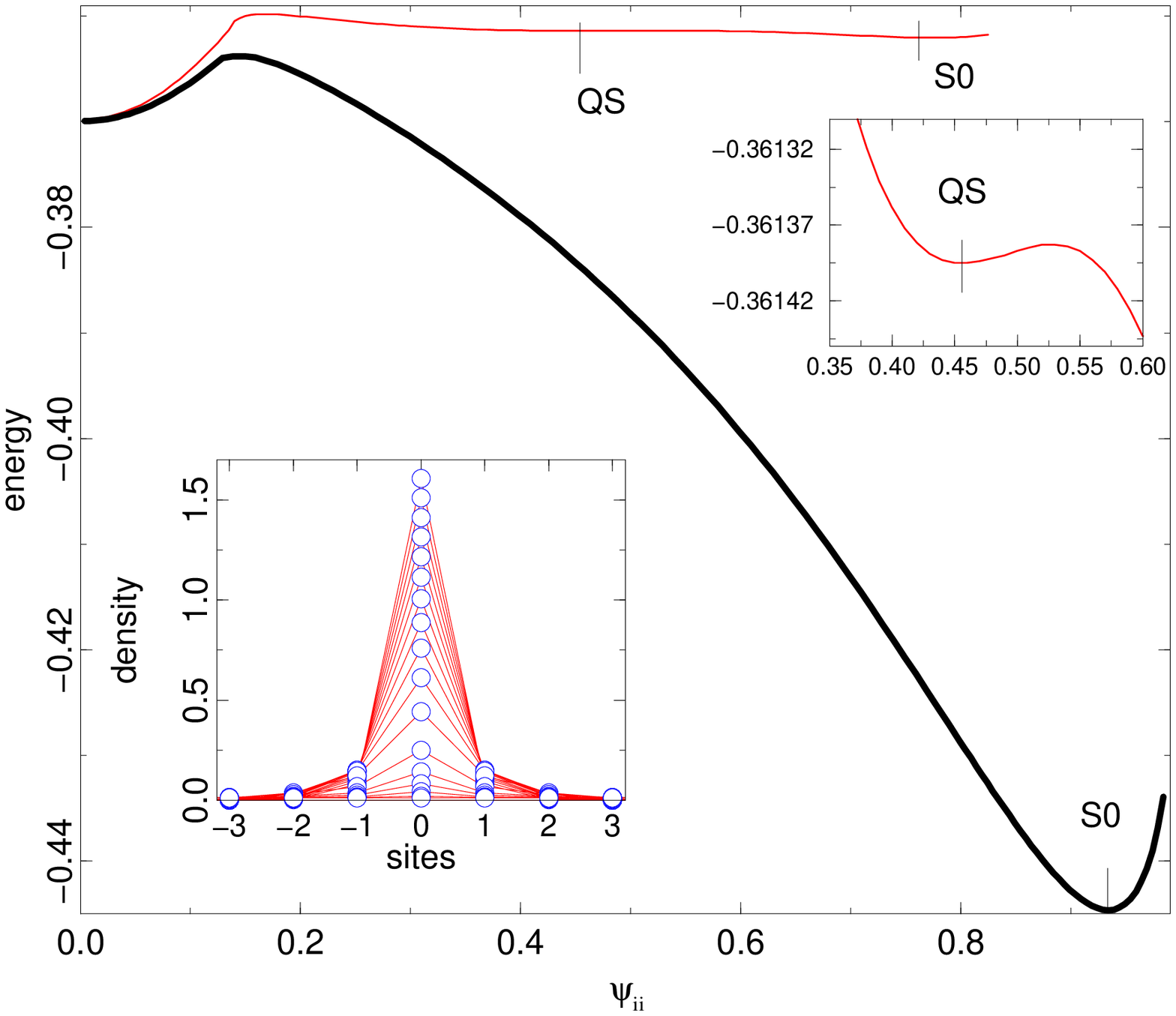}
\caption{\label{fig:10} Top: Radial PN energy barrier versus $\psi_{i,i}=c$ 
between (S0) and (QS) at  $t=0.08$, $U=0.23$ Insert: Profile variation 
of the bipolaron electronic density  along the same path. 
Bottom: Same at $t=0.092$, $U=0.1$ between bipolaron (S0) and the 
extended state at $\psi_{i,i}=0$ (full line)
and $t=0.092$,$U=0.204$ between bipolaron (S0), (QS) as an 
intermediate state and the extended state (thin line).
Insert: Bipolaron profile (bottom left) and magnification of the (QS) region
(top right).}
\end{center}\end{figure}

It is also worthwhile to note that there is also a PN energy barrier between
the bipolaron (S0) and (QS), 
which have the  same symmetry (see fig.\ref{fig:10}). We tested that it 
does not generate any path with a lower PN energy barrier when shifting the bipolaron (S0) or (QS)
by one lattice spacing. For that, we compare the energy barrier obtained 
for shifting (S0) by one lattice spacing via the direct path (S1) $E_{B}(S0 \rightarrow S1)$, 
to that obtained by
the indirect path (S0) via (QS) and (S1) involving the jump of two consecutive barriers
$E_{B}(S0 \rightarrow QS)$ and $E_{B}(QS \rightarrow S1)$. We found that the energy barrier 
between (S0) and (QS) was always relatively too high to favor 
the indirect path.

\begin{figure} \begin{center}
\includegraphics{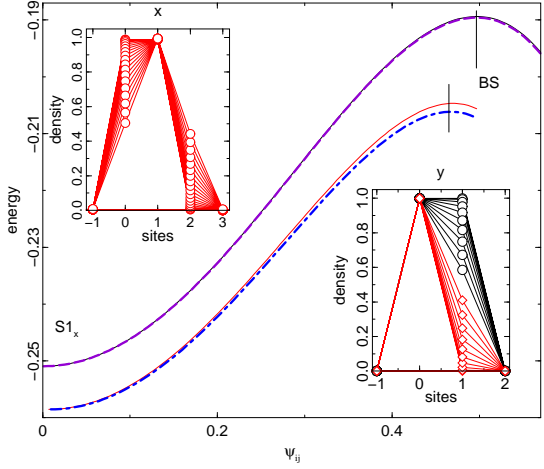}
\caption{\label{fig:11}  Same as fig.\ref{fig:7} but for bipolaron (S1) 
moving in the longitudinal direction with the constraint $\psi_{j,k}=c$
or rotating transversally with the constraint $\psi_{i,l}=c$ at 
 $t=0.01$, $U=0.3$ (upper lines), $t=0.03$, $U=0.3$ (lower lines).
(Vertical lines indicates the location of the energy extrema 
corresponding to bipolaron (S1) and (BS) or (BS$^{\prime})$).
Although almost the same, the curves for the transversal 
motion are slightly lower than those for the longitudinal motion.
Insert: Variation of the Profile of electronic density along 
the axis $i-j$ of bipolaron (S1) for the two 
continuous  paths of the bipolaron  at $t=0.03$, $u=0.3$
corresponding to the longitudinal motion (left top) and the 
transversal motion (right bottom).}
\end{center}\end{figure}

When bipolaron (S1) becomes the ground-state, there are two PN energy barriers depending on the
direction it is displaced, transversally or longitudinally.
If it is displaced longitudinally in the direction of the bond $(i,j)$ 
where (S1) is localized, the minimax may be obtained by
varying the constraint $\psi_{j,k}=\psi_{k,j}=c$ which tends to displace (S1) longitudinally.
Fig.\ref{fig:11} shows this energy variation versus $c$ starting from bipolaron (S1).
The maximum along this path corresponds to the longitudinal star bisinglet 
bipolaron (BS). For $t=0.03$ and $U>0.28$, it costs less energy to use this path with 
minimax  (BS$^{\prime}$) than to use the path with minimax (S1/S0) passing by
bipolaron (S0). 

The transversal PN energy barrier of bipolaron (S1) can be also calculated.
Actually, the transversal motion of (S1) with the lowest PN energy 
barrier has to be done in two steps (in the anti-integrable limit).
If we denote by $(i,j,k,l)$ the corner sites of an elementary square of the 2D lattice
and move (S1) from the bond $i-j$ to bond $l-k$, then (S1) rotates 
once by $\pi/2$ around the center site $i$ and again
by $\pi/2$ but around the center site $l$. These two jumps have the 
same PN energy barrier. It can be measured from the height of the minimax
determined by the constraint $\psi_{i,l}=\psi_{l,k}=c$. This path yields 
the $2$ star multisinglet (BS$^{\prime}$) with a diagonal symmetry 
axis where $\psi_{i,j}=\psi_{i,l}$ and where the two branches $(i,j)$ and $(i,l)$ 
are orthogonal. It is found that the longitudinal and transverse PN 
energy barrier are almost equal.

\begin{figure} \begin{center}
\includegraphics{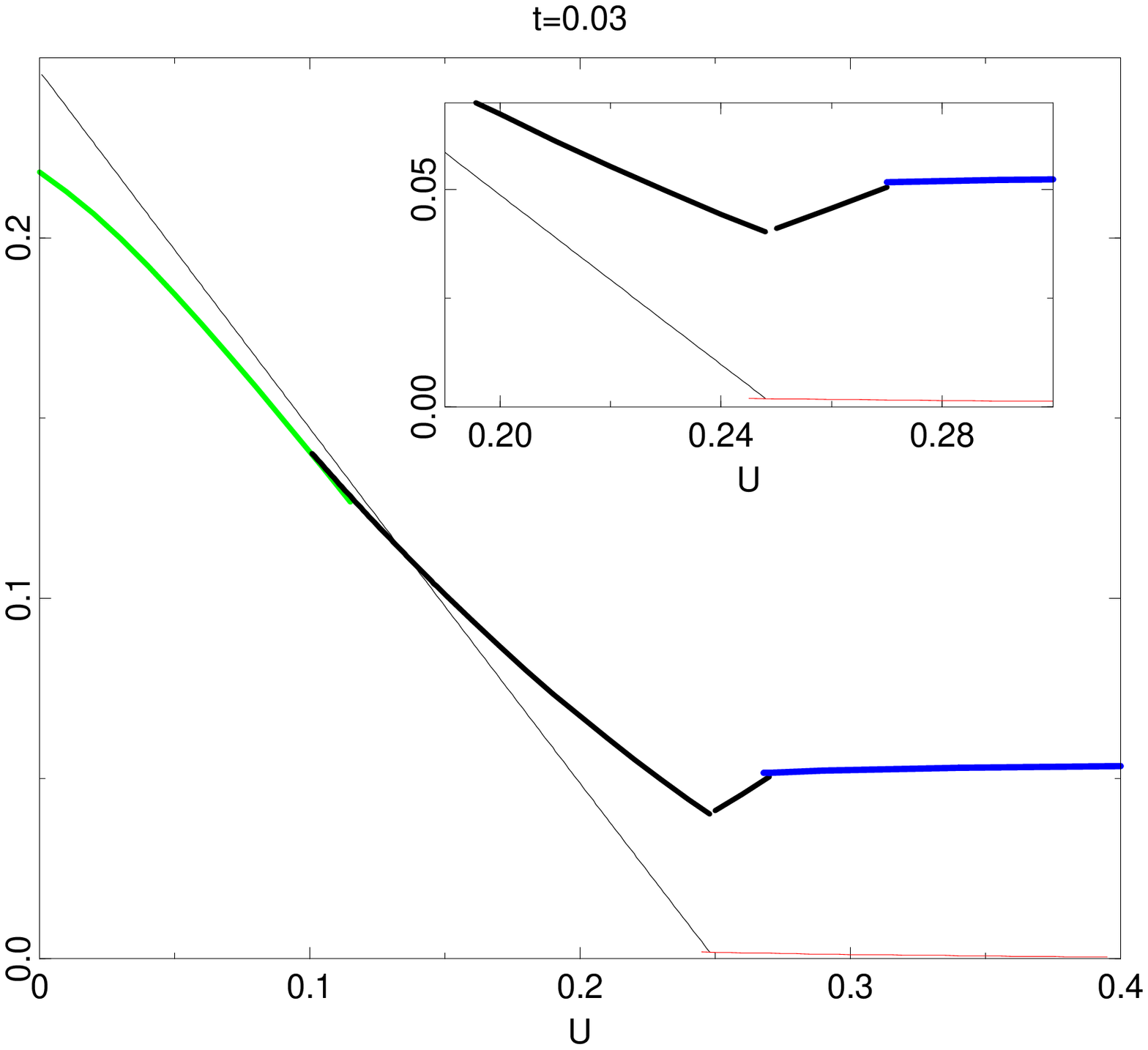}
\includegraphics{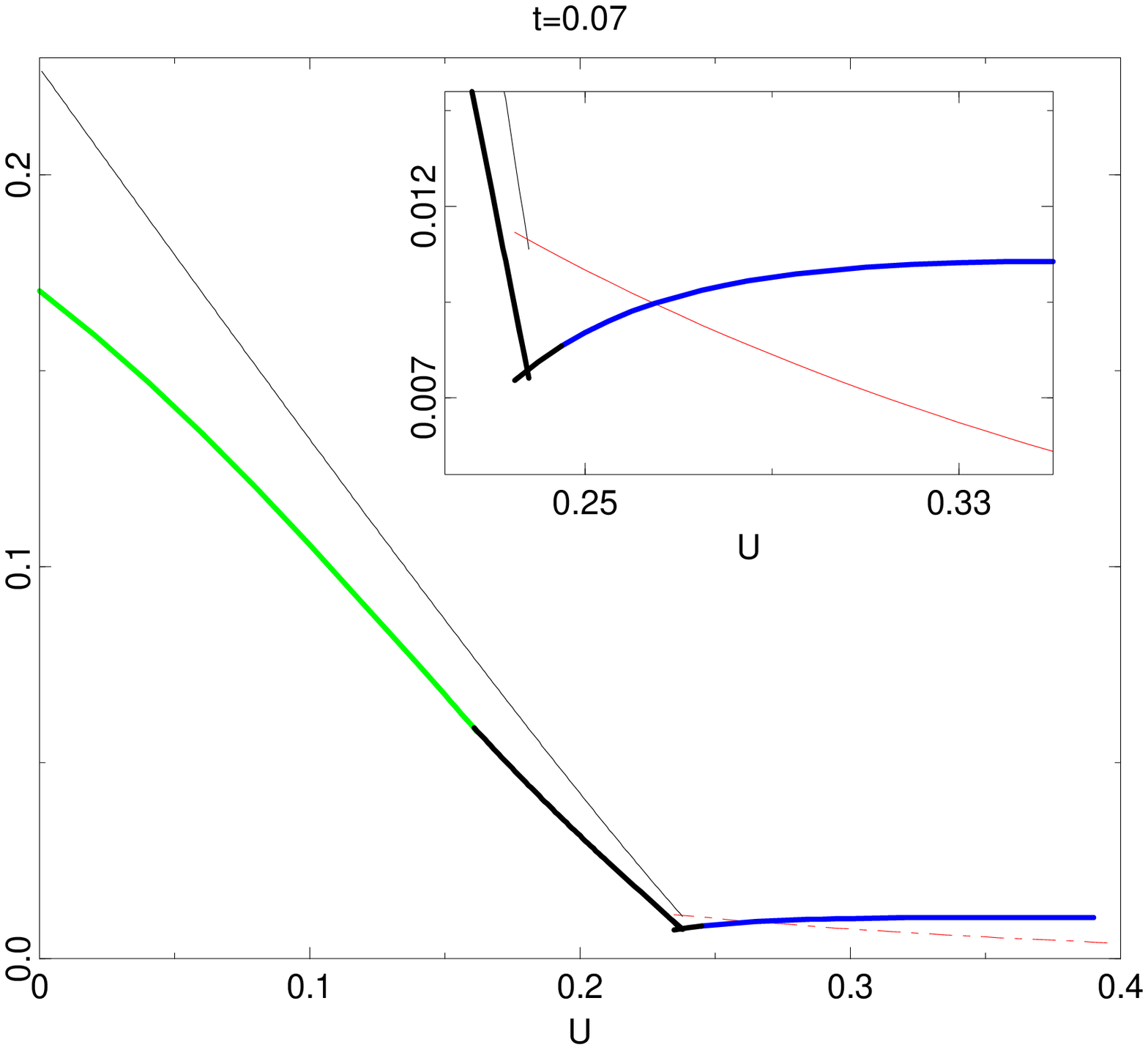}
\includegraphics{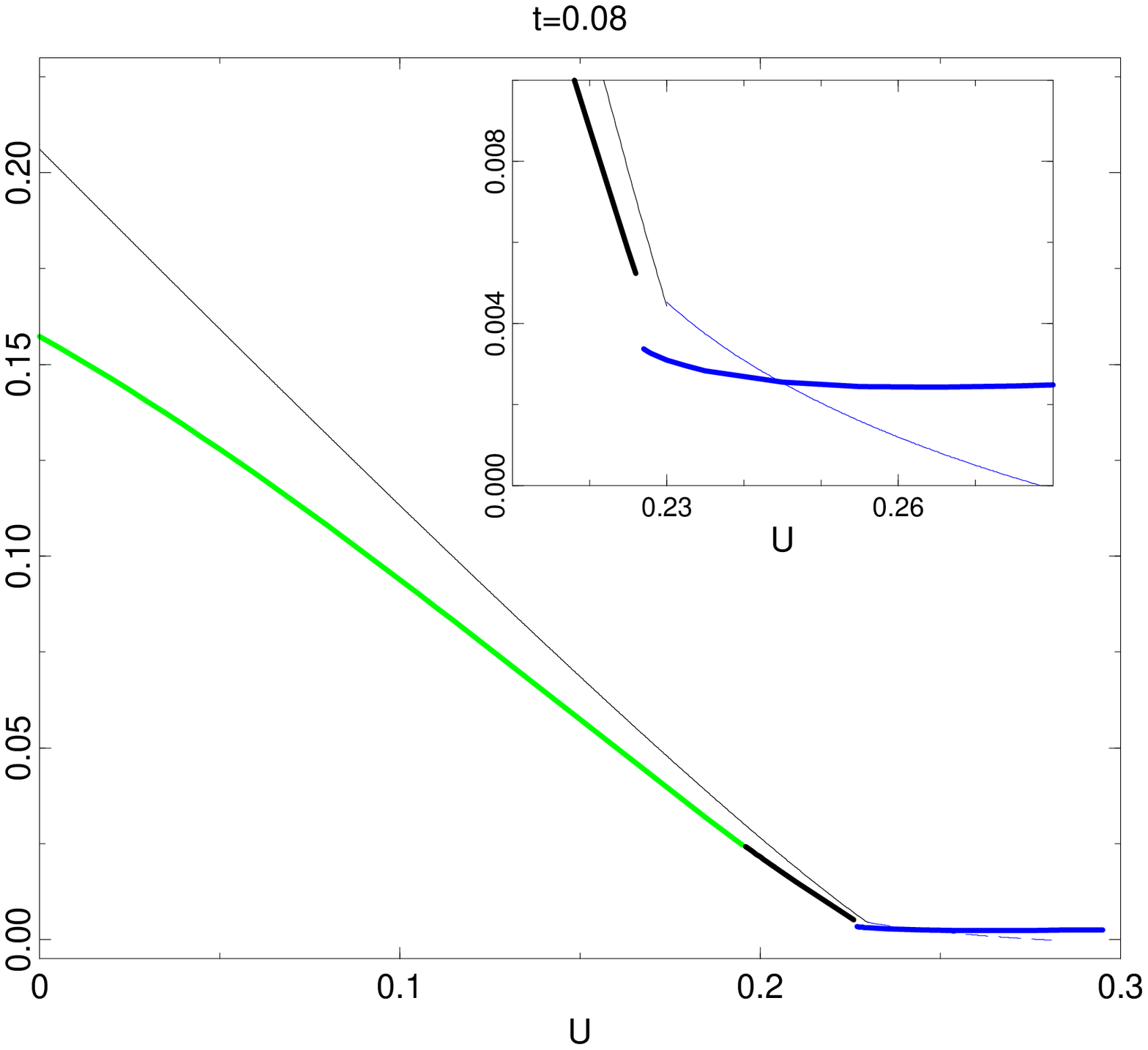}
\caption{\label{fig:12} Peierls-Nabarro energy barrier (thick lines)
and Binding energy (thin lines) of ground-state bipolarons  versus $U$ 
for different values of $t$: $t=0.03$ (top), $t=0.07$ (middle) and
$t=0.08$ (bottom). The lines are broken because of the change of 
minimax. Inserts: magnification in the region of
first order transition.}
\end{center}\end{figure}

The precise determination of the PN energy barrier becomes more 
delicate close to the border line of the phase diagram fig.\ref{fig:2}
with the domain of extended electrons. Here the binding energy of 
the bipolaron becomes very weak. To move a bipolaron, it may cost less energy to pass
the energy barrier for breaking the pair of electrons (fig.\ref{fig:10}), 
producing extended electronic states, and next to pass a new equivalent
energy barrier for reconstructing the bipolaron at another site.

Figs.\ref{fig:12} gather  the resulting PN energy barrier versus $U$ and for several values
of $t$ obtained by comparison of these different paths (for that reason we have
broken lines with possible discontinuities). The essential result is that
close to the region of the triple point between bipolaron (S0), (S1) and (QS), the 
Peierls-Nabarro energy barrier sharply drops and reaches the same order of 
magnitude as the binding energy of the bipolarons. When $U$ slightly
increases beyond this point, the binding energy of the bipolaron sharply decreases. 
Conversely, when $U$ slightly decreases, the PN energy barrier sharply increases.
In that region,  the paths allowing a shift by one lattice spacing
of a bipolaron with the smallest PN energy barrier involves the 
successive transformations \ldots (QS) $\to$ (S1) $\to$ (QS) \ldots. 

There are regions in the parameter space where the 
bipolaron binding energy becomes very small, for example when $t$ is small and $U>1/4$)
(e.g. fig.\ref{fig:12} at $t=0.03$).
The PN energy barrier of a bipolaron is then practically equal to the PN energy barrier of a 
free single polaron. In other regions, close to the first-order transition border line 
with the extended states, the  bipolaron binding energy also becomes very small,
but then the PN energy barrier is practically that which has to be 
overcome for the electron delocalization.

\section{Concluding remarks}

The interplay between the electron-phonon coupling and the direct electronic
repulsion has been treated accurately in the adiabatic Holstein-Hubbard model in
two dimensions. Numerical investigations complemented by analytic variational 
calculations yield the phase diagram of the ground-state of a single 
bipolaron, which consists of several domains separated by first order transition 
lines (see fig.\ref{fig:2}).  It is found that the different bipolaronic states
that are obtained, already exist at the  anti-integrable limit and can be generated from this
limit by  continuation.

There is an interesting region in the phase diagram where the 
bipolaronic ground-state becomes  a quadrisinglet bipolaron, which is
a superposition of four singlets sharing one central site.
The binding energy of that bipolaron is the result of the spin resonance
between a strongly localized polaron and a peripheral electron localized on its 
nearest neighbors.

There is a triple point where the three kinds of bipolaron coexist with 
the same binding energy, which is still significantly large and non-negligible.
The internal modes of the bipolarons soften significantly in that 
region. Moreover, the Peierls-Nabarro energy barrier (PN) of the bipolaron
in that region is strongly depressed, which improves the classical mobility 
of this bipolaron. This effect is related to the appearance of several 
intermediate metastable bipolaronic state which have almost the same energy.
A small variation of parameters $(U,t)$ in that region suffices 
either to lift the near degeneracy, with a PN energy barrier which 
grows very fast, or to depress sharply  the binding energy of the bipolaron itself.
The energy landscape around the bipolaron has been explored.
It has been found that it is quite flat in the region of the triple 
point with several minimum energy states close in energy and small energy barriers 
between them.

These features strongly support the conjecture that the quantum tunneling of the bipolaron will be 
strongly enhanced in the vicinity of this triple point due both to the 
small PN energy barrier and to the hybridization between the nearly degenerate states.
This assertion will be confirmed by the results of the next paper 
where the quantum lattice fluctuations will be treated as perturbation
through a tight binding model \cite{PA99}. 

Unlike the conclusion of ref.\cite{CRF98}, we find a plausible mechanism 
for a drastic reduction, under specific conditions by several orders of magnitude, of
the effective mass  of a bipolaron while preserving a relatively large binding energy.
Let us recall that fig.\ref{fig:4} shows that the binding energy close to the triple point 
is still about $0.005 E_{0}$. Since $E_{0}=8 g^{2}/\hbar \omega_{0}$ could 
reach in some realistic physical models a magnitude of about $10 eV$, this binding
energy can be close to $0.05 eV$ which corresponds in temperature 
units to  $500$ K!. In the same region, the Peierls-Nabarro energy 
barrier has a value almost equal to this binding energy. It is drastically 
reduced compared to what could be expected for small bipolarons in 
standard theories. Note that is about $50$ times smaller than the Peierls-Nabarro 
energy barrier of the bipolaron (S0) at $U=0$ and the same value of 
$t$.  

When the temperature of the system goes below this characteristic 
temperature where bipolarons can form,
and if the tunnelling energy could reach comparable values (which will be 
shown in the next paper), this effect might be sufficient
to favor a superfluid state at 0K against either a bipolaron ordering
or a magnetic polaron ordering. This state could persist to unusually 
large temperatures. There is, of course, another condition, 
which is that the direct bipolaron interactions are not too strong. 
This is very unlikely to be true at half-filling, where the polarons are 
close-packed, but this condition might become fulfilled when the 
density of electrons moves sufficiently far from half-filling. 
These quantitative results in the Holstein-Hubbard model yield a more quantitative support to
earlier but less specific conjectures that high Tc superconductivity could be explained
by a well balanced competition between electron-electron repulsion and electron-phonon 
interaction \cite{Aub93b,Aub95b}.

The methods used above should also work with other perturbations from the
anti-integrable limit. In the present paper, the Laplacian form for the kinetic energy
implies that the bipolaron ground-state when it has the square symmetry, 
has necessarily the trivial quantum symmetry $(s)$.
However, it is not physically unrealistic to assume that the electronic kinetic energy terms
in Hamiltonian  (\ref{hamiltonian})  might be different from a discrete Laplacian form
\footnote{A reduced Holstein-Hubbard Hamiltonian on the copper square 2D sublattice of 
cuprates should involve more than nearest-neighbor transfer integral 
due to the oxygen bridges.}.
When there are second-nearest-neighbor electronic transfers with  
significant amplitudes (but not necessarily as large as the nearest-neighbor integral) and
with appropriate signs, the ground-state electronic wave function
$\{\psi_{i,j}\}$ of bipolaron (QS) should have a $(d)$ symmetry (see 
appendix \ref{AppC}). 
A superfluid state of such bipolarons with a degenerate internal quantum symmetry
could be perhaps  related with the now well accepted fact that the superconducting
order parameter of cuprates has a (d) wave symmetry \cite{Pin97}. Further 
works will investigate consequences of this quantum symmetry. We expect that 
when the (d) wave symmetry is favored by appropriate terms, the 
stability domain of the bipolarons that can take advantage of this 
symmetry will be extended: that is, those of bipolaron (QS).

In principle, the method used in this paper for calculating adiabatic bipolarons
could be extended to more complex and realistic models. There are many kinds of bipolarons
in the anti-integrable limit, as shown in appendix (\ref{AppA}). It is 
not obvious that only bipolarons (S0), (S1) or (QS) are competing 
as ground-states. More generally, if there are more transfer 
integrals between further neighbors, other $N_{2}$ star multisinglets 
with $N_{2}>4$ (eg. $N_{2}=8$ etc\ldots) might become more favorable
as bipolaron ground-state in some cases.

Of course, the present study with only two electrons is by far not 
sufficient to describe real cuprates where  the density of 
electrons is close to half filling. Our model suffers in that we are not 
yet able to describe in a satisfactory fashion the interactions 
between the bipolarons.  Nevertheless, we believe that we already obtained
useful informations on the effect of competing strong electron-phonon and 
strong electron-electron interactions. Our approach supports the possibility of a
bipolaronic mechanism to explain high Tc cuprate superconductors, and this might be a
clue for a more consistent explanation for the origin 
of high $T_c$ superconductivity in real high Tc superconducting cuprates. 

Of course, one may argue against our approach that assuming a large tunnelling 
energy for bipolarons is a warning that the system might not be well described 
anymore by perturbative methods from the adiabatic limit. But, our results also warn that 
perturbative methods from a Fermi liquid model with strong electron
interactions is also quite far from its limit of validity because of 
the non-negligible lattice
distortions that could be generated. 
From a strict mathematical point of view, there are no reasons why 
the same physical state could be not described from different limits, so 
that a debate about this question of principle is useless. In the 
end, only the efficiency and 
simplicity of a theory, are the right criteria for physicists.

We thank R.S. MacKay and C. Baesens for useful discussions.  One of us (LP) acknowledges DAMTP
in Cambridge for its hospitality during the completion of this manuscript.

\appendix

\section{Bipolaron States at the Anti-integrable Limit}
\label{AppA}
We describe an elementary classification of the bipolaron solutions in the
anti-integrable limit. It could also probably be obtained as a special case of two electrons
of the more sophisticated  homology theory that was recently developed by Baesens
and MacKay \cite{BMc98} for the pure Holstein model with many electrons.

These bipolaron solutions $\{\psi_{i,j}\}$ fulfill eq.\ref{eleigen} and 
\ref{NLEQ} at $t=0$,  which yields

\begin{equation}
\left(-\frac{1}{4} (\rho_i+\rho_j) + U \delta_{i,j}\right) \psi_{i,j}=
F_{el}  \psi_{i,j}\label{eleigenai}
\end{equation}
where the electronic density $\rho_i$ is defined by eq.\ref{rho_i}. 

We first note that for any solution of this equation, the phases of the complex numbers
$\psi_{i,j}$ can be chosen arbitrarily and independently. Thus, in 
this paper, we remove this trivial degeneracy by choosing their phases to be zero,
that is $\psi_{i,j}$ is assumed to be real positive. However, it 
could be removed by fixing another symmetry for the bipolaron (e.g. symmetry 
$(s^{\prime})$ or $(d)$). In principle, removing the phase degeneracy is necessary
to allow a unique continuation of a solution  at $t \neq 0$ 
(if there is no other continuous degeneracy). Since we noted that the bipolaron ground-state
at $t\neq 0$ necessarily fulfills this condition, it could be found among
these continued solutions.  Actually, this trick is analogous to that used in 
ref.\cite{MA94} for proving the existence of discrete breathers.

There is another trivial degeneracy at $t=0$  but that is now discrete.
Any solution  of eq.\ref{eleigenai} yields infinitely many other solutions
with the same energy, which are simply obtained by arbitrary
permutations of the sites  of the lattice $j=\mathcal{P}(i)$.

For each solution, the set of occupied sites $i \in S$ is defined by
the condition $\rho_i \neq 0$. We call \textit{link} a pair of sites $(i,j)$
such that $\psi_{i,j} \neq 0$ ( which implies $i \in S$ and $j \in S$). 
A bipolaron state at $t=0$ is said to be connected if the graph generated
by all the links is connected.

We first investigate the connected states of eq.\ref{eleigenai}, which implies that
when $\psi_{i,j} \neq 0$,

\begin{equation}
U \delta_{i,j}-\frac{1}{4}(\rho_i+\rho_j)=F_{el}.
\label{Cstpot}
\end{equation}
$F_{el}$ is independent of the pair of connected sites $(i,j)$. Considering
two different sites $i$ and $j$ connected to a third site $n$,
it comes out that $\rho_i+\rho_n=\rho_j+\rho_n$, which implies $\rho_i=\rho_j$.
More generally, two occupied sites connected by some path with an even
number of links have necessarily the same electronic density. As a
result, the set of occupied sites $S$ is the union of two disjoint sets
of sites $S_1$ and $S_2$ where the electronic densities are the same.
For  $i \in S_1$, the electronic density is $\rho_i=\rho_1$ and 
for $j \in S_2$ , $\rho_j=\rho_2$. Moreover, sites $i \in S_1$ are only linked to
sites $j \in S_2$ and vice versa.

We consider separately the connected bipolaron states without and with
doubly occupied sites $i$. These sites  are defined by the condition $\psi_{i,i} \neq 0$.

\subsection{Connected Bipolaron States with no Doubly Occupied Site} 
\label{AppA2}

If $\psi_{i,i}=0$ for any $i$, the electronic wave function is
a normalized combination of two sites singlet states defined in eq.\ref{bip1}, and
consequently  these states and their energies do not depend on $U$.

$N_s=N_{1}+N_{2}$ is defined as the total number of occupied sites, $N_1$ is the number
of sites in $S_1$ with density $\rho_1$ and $N_2$ the number of sites 
in $S_2$ with density $\rho_2$. Since the total number of electrons 
is two,

\begin{equation}
N_1 \rho_1 + N_2 \rho_2 = 2 .
\label{globdens}
\end{equation}

It follows from eq.\ref{Varform} that
\begin{equation}
F_v=-\frac{1}{8}  \sum_i \rho_i^2=
-\frac{1}{8} (N_1 \rho_1^2 + N_2 \rho_2^2)
\label{Var2sit}
\end{equation}
and equivalently from eq.\ref{Varform2} and eq.\ref{Cstpot}
\begin{equation}
F_v= -\frac{1}{4}(\rho_1+\rho_2)+\frac{1}{8} (N_1 \rho_1^2 + N_2 \rho_2^2)
\label{Var2sit2}
\end{equation}

Identifying the two results (\ref{Var2sit}) and (\ref{Var2sit2}), and
using eq.\ref{globdens}, two solutions come out which are first

\begin{eqnarray}
\rho_1&=&\frac{1}{N_1} \quad \mbox{and} \quad \rho_2=\frac{1}{N_2} 
\quad \mbox{with} \label{sol1}\\
F_v&=&-\frac{N_s}{8N_1N_2} \label{en1}
\end{eqnarray}
and second
\begin{eqnarray}
\rho_1&=&\rho_2=\frac{2}{N_{s}} \quad \mbox{with} \label{sol2}\\
F_v&=&-\frac{1}{2N_s} \label{en2}
\end{eqnarray}

In the first case(\ref{sol1}), we have $\rho_1 \neq \rho_2$ when $N_1 \neq N_2$.
Then $\psi_{i,j} \neq 0$ when $i \in S_1$ and $j \in S_2$ or vice versa. This condition
determines a rectangular $N_1 \times N_2$ matrix. The square of its real positive coefficients
fulfills  the linear eqs.\ref{rho_i}

\begin{eqnarray}
\sum_{j \in S_2} \psi_{i,j}^2 &=& \frac{1}{2 N_1} \label{sumline} \\
\sum_{i \in S_1} \psi_{i,j}^2 &=& \frac{1}{2 N_2} \label{sumcol}
\end{eqnarray}

A particular solution of this set of equations is $\psi_{i,j}^2=1/(2 N_1 N_2)$. However,
there are $N_1+N_2$ linear equations to determine the $N_1 N_2$ coefficients.
Except for the case $N_1=1$ or equivalently $N_2=1$, (we have 
$N_{1} \neq N_{2}$), this solution  $\psi_{i,j}$ is
degenerate and belongs to a nonvoid bounded and compact domain defined by
the positivity of $\psi_{i,j}^2$.

\begin{figure} 
\begin{center}
\includegraphics{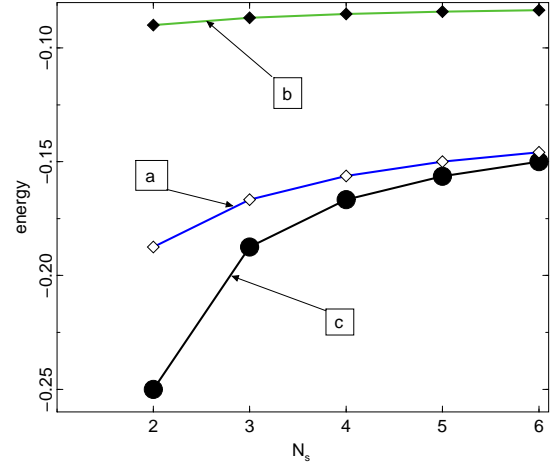}
\caption{\label{fig:13}  At $t=0$, energy as a function of $N_{s}$:
(a) star sisters solutions for $U=0.25$,
(b) star sisters solutions $U=0.4$, and
(c) star multisinglets (S1) for $N_{s}=2$, bisinglet (BS)
for $N_{s}=3$, trisinglet (TS) for $N_{s}=4$, quadrisinglet 
(QS) for $N_{s}=5$.}
\end{center}
\end{figure}

The solutions with $N_1=1$ appear especially interesting, not only because they are not
continuously degenerate but because their energy $F_v=-(N_2+1)/(8N_2)<-1/8$ is
significantly lower than zero. It is not far above those of the bipolaron (S0) which
is $F_v=-1/2+U$ and those of the 
singlet bipolaron (S1) which is $F_v=-1/4$ (see fig.\ref{fig:13}).
We call them star multisinglets. (S1) and (QS) are star multisinglets with $N_2=1$ and $N_2=4$
(see fig.\ref{fig:1}).

In the second case (\ref{sol2}) or in the first case when $N_1=N_2$, the
electronic densities at the $N_s=2N_{1}$ occupied sites are equal.
Then there are in general $N_s(N_s-1)/2$ nonzero coefficients $\psi_{i,j}=\psi_{j,i}$ 
since there are no doubly occupied sites ($\psi_{i,i}=0$).  They fulfill $N_s$
equations $\sum_j \psi_{i,j}^2 = 1/(2N_s)$ (\ref{rho_i}).
Again, this system has a trivial solution, which is $\psi_{i,j}^2=1/(2N_s(N_s-1))$
for $i \neq j \in S$. However, when $N_s \geq 4$, this system of equations
becomes underdetermined and yields continuously degenerate solutions
that belong to a compact domain since again $\psi_{i,j}^2$ must be 
found positive.

It is worthwhile to remark that although these solutions were assumed to be 
connected, when they form a continuously degenerate set this set
may contain non-connected states just at the border of the compact domain of solutions.
For example, in this second case, let us split the set $S$ of occupied sites
in two disjoint subsets $T_1$ and $T_2$ with $M_s \geq 2$ and $N_s-M_s\geq 2$ sites respectively.
Let us set $\psi_{k,l}=\psi_{l,k}=0$ for $k \in T_1$ and $l \in T_2$, which corresponds to
$M_s(N_s-M_s)$ conditions. When $N_s(N_s-1)/2 -M_s(N_s-M_s) \geq N_s$, the equation 
$\sum_j \psi_{i,j}^2 = 1/N_s$ has the solution  $\psi_{i,j}=1/\sqrt{(M_s-1)N_s}$
for $i \neq j \in T_1$ and $\psi_{i,j}=1/\sqrt{(N_s-M_s-1)N_s}$ for $i \neq j \in T_2$.
This situation is found to occur for $N_s \geq 6$.

\subsection{Connected Bipolaron States with
Doubly Occupied Sites at $U \neq 0$}
\label{AppA3}

Let us require $N_s>1$ to avoid the onsite bipolaron (S0).
Such connected solutions $\{\psi_{i,j}\}$ of
eq.\ref{eleigenai} have at least one doubly-occupied site $k$ ( i.e such that
$\psi_{k,k} \neq 0$). The set of occupied sites can be split
in two sets $S_1$ and $S_2$ with electronic density $\rho_1$ and 
$\rho_2$, respectively.  Let us consider a doubly occupied site $k$ which we assume
to belong to the set of sites $S_1$ with electronic density $\rho_1$.
Let us also consider a site $l \in S_2$ such that $\psi_{k,l}\neq 0$. There
exists such a site since the solution is connected. Then, because
of eq.\ref{Cstpot}, we have
\begin{equation}
F_{el}= -\frac{1}{2} \rho_k+U=-\frac{1}{4} (\rho_k+\rho_l)=
-\frac{1}{4}(\rho_1+\rho_2)=-\frac{1}{2} \rho_1+U
\label{rhodenst}
\end{equation}
which implies 
\begin{equation}
	U=\frac{\rho_1-\rho_2}{4}
	\label{Uvalue}
\end{equation}

All the doubly occupied sites must belong to $S_1$. The nonzero Hubbard amplitude
$U$ obviously implies distinct electronic density for doubly-occupied
and non-doubly-occupied sites. It follows from eq.\ref{globdens} that
\begin{equation}
\rho_1=2 \frac{1+2 U N_2}{N_s} \quad \mbox{and} \quad
\rho_2=2 \frac{1-2 U N_1}{N_s} \label{rho_U}
\end{equation}
which implies
\begin{equation}
	-\frac{1}{2N_{2}}<U<\frac{1}{2N_{1}}
	\label{cdens}
\end{equation}
in order that both $\rho_{1}$ and $\rho_{2}$ be positive.

We now obtain from eqs.\ref{Varform},\ref{Varform2} at $t=0$
\begin{equation}
F_v=\frac{1}{8} (N_1 \rho_1^2 + N_2 \rho_2^2)-\frac{1}{4}(\rho_1+\rho_2)
\label{energv}
\end{equation}
and substituting \ref{rho_U} in \ref{energv}
\begin{equation}
F_v=\frac{1}{2N_s}[4U^2N_1N_2+2U(N_1-N_2)-1]
\label{energv2}
\end{equation}

According to eq.\ref{rho_i}, the set of electronic states $\psi_{i,j}=\psi_{j,i}$ satisfy 
\begin{eqnarray}
	 \frac{\rho_1}{2} & = & \psi_{i,i}^{2}+\sum_{j\in S_2} \psi_{i,j}^2 
	 \quad \mbox{for} \quad i \in S_{1}
	\label{rhoeq1}  \\
     \frac{\rho_2}{2} & = & \sum_{i\in S_1} \psi_{i,j}^2  \quad \mbox{for} 
     \quad j \in S_{2}
	\label{rhoeq2}
\end{eqnarray}
where $\rho_{1}$ and $\rho_{2}$ are given by eqs.\ref{rho_U}.
There are $N_1+N_2$ linear equations for calculating $N_1(N_2+1)$ 
positive numbers $\psi_{i,i}^{2}$ with $i \in S_{1}$ and 
$\psi_{i,j}^{2}=\psi_{j,i}^{2}$ with $i \in S_{1}$ 
and $j \in S_{2}$. A particular solution is 
obtained when all $\psi_{i,j}^{2}=\rho_{2}/(2N_{1})= (1-2UN_{1})/(N_{s}N_{1})$ and
all  $\psi_{i,i}^{2}=(N_{1}\rho_{1}-N_{2}\rho_{2})/(2N_{1})=(N_{1}-N_{2}+4 
U N_{1}N_{2})/(N_{1}N_{2})$.
The positivity of $\psi_{i,i}^{2}$ requires 
\begin{equation}
 \frac{N_{2}-N_{1}}{4N_{1}N_{2}} < U	
	\label{condpos}
\end{equation}
Actually, there are no  positive solutions at all to eqs.\ref{rhoeq1} and 
\ref{rhoeq2} when this condition is not fulfilled.

When conditions (\ref{cdens}) and (\ref{condpos}) are fulfilled and when $N_{1}>1$, the number of 
variables exceeds the number of equations, and there is a compact set
of degenerate positive solutions to the linear eqs.\ref{rhoeq1} and \ref{rhoeq2}. 

When $N_{1}=1$ and  when
\begin{equation}
	\frac{N_{2}-1}{4N_{2}}\leq U \leq \frac{1}{2}
	\label{condsist}
\end{equation}
there is a unique solution to this set of linear equations. For a given $N_s$,
it has the lowest energy $F_v$ (plotted on fig.\ref{fig:13} for different values of $U$).
This solution is called $N_2$ star sister. It can be interpreted as a mixing between
the bipolaron (S0) and the $N_2$ star multisinglet. We denote (S0/S1) the one star sister
that mixes both (S0) and (S1) etc\ldots.
According to the implicit function theorem, this nondegenerate solution can be continued
to $t$ nonzero except at the bifurcation points at $U=1/2$ and 
$U=(N_2-1)/ 4N_2$
where this solution bifurcates with (S0) and the $N_2$ star multisinglet, respectively.
In the anti-integrable limit $t=0$, its energy is larger than both those of (S0) 
and $N_2$ multisinglet (see part \ref{PNpart} for $t>0$).

\subsection{Non Connected Bipolaron States} \label{AppA4}

Let us now assume that we have a solution $\{\psi_{i,j}\}$ of  eq.\ref{eleigenai}, which is 
not connected. Then, it can be  decomposed into a sum of 
normalized connected components $\{\psi_{i,j}^{\alpha}\}$.
We define $S_{\alpha}$ as a set of lattice sites that are 
connected  with each other by a sequence of links. 
$\psi_{i,j}^{\alpha} \neq 0$ is proportional to $\psi_{i,j}$ for $i \in 
S_{\alpha}$ and $j \in S_{\alpha}$ and zero elsewhere. The 
proportionality coefficient $\lambda_{\alpha}$ is chosen  in order 
that $\{\psi_{i,j}^{\alpha}\}$ normalized. Then we have 

\begin{equation}
	\psi_{i,j} = \sum_{\alpha} \lambda_{\alpha}\psi_{i,j}^{\alpha}
	\label{decompdis}
\end{equation}
with 
\begin{equation}
\sum_{\alpha} |\lambda_{\alpha}|^{2} =1	
	\label{normc}
\end{equation}
Two components $\alpha$ and $\beta$ have no common occupied site.

Then, $\{\psi_{i,j}^{\alpha}\}$ is a connected solution of eq.\ref{eleigenai},
for the eigenenergy  $F_{el}^{\alpha}= F_{el}/|\lambda_{\alpha}|^{2}$ and 
the Hubbard term $U_{\alpha}= U/|\lambda_{\alpha}|^{2}$.

If component $\alpha$ has no doubly occupied sites, it does not 
depend on $U$. Then if $N_{1}^{\alpha}+N_{2}^{\alpha}=N_{s}^{\alpha}$ represents
the number of sites in each of the two groups of sites with different 
electronic densities defined at the beginning of this appendix,
eq.\ref{Cstpot} implies

\begin{equation}
	|\lambda_{\alpha}|^{2} =  - 4 	F_{el}  
 	\frac{N_{1}^{\alpha}N_{2}^{\alpha}}{N_{s}^{\alpha}}
 	\label{undoccs}
\end{equation}

If component $\alpha$ has doubly occupied sites (see appendix. \ref{AppA3}),
then  eq.\ref {eleigenai} implies

\begin{equation}
 	|\lambda_{\alpha}|^{2} =  - N_{s}^{\alpha} F_{el} 
 	-U(N_{2}^{\alpha}-N_{1}^{\alpha})
 	\label{doccs}
\end{equation}

There are constraints for solving the second equation because of 
inequalities (\ref{cdens}) and (\ref{condpos}), which imply

\begin{eqnarray}
	 - \frac{1}{2N_{2}^{\alpha}} & < &\frac{U}{|\lambda_{\alpha}|^{2}}< 
	 \frac{1}{2N_{1}^{\alpha}} 
	\label{csts1}  \\
	\frac{N_{2}^{\alpha}-N_{1}^{\alpha}}{4N_{1}^{\alpha}N_{2}^{\alpha}} & < & 
	\frac{U}{|\lambda_{\alpha}|^{2}}
	\label{csts2}
\end{eqnarray}

Conversely,  we can construct  non-connected
bipolaron states which are combination of non-overlapping connected states. 
Then the amplitude  $|\lambda_{\alpha}|^{2}$ is defined by eqs. \ref{undoccs}
or \ref {doccs}, but then we have to choose $F_{el}$ in order that the 
normalization condition (\ref{normc}) is fulfilled.
This is easy to do when only components $\alpha$  with no doubly occupied sites
are involved. Otherwise, we have to take into account the constraints 
(\ref{cdens}) and (\ref{condpos}). There are many such solutions but we 
did not investigate them in detail. Some of them are easy to find: for 
example, 
the nonconnected solution with two components involving the bipolaron (S0) 
located on two adjacent sites. This solution is called (2S0)
and has the energy $U-1/4$.

It can be checked that the energy of the disconnected state is always larger than those of
its components with the smallest energy. 

\section{Two-site model} \label{AppB}

It is instructive to analyze all the extrema of the variational form (\ref{Varform}) 
on a  lattice reduced to only two sites $i$ and $j$, because it can 
be explicitly calculated in all detail. However, a limitation of this 
restricted model is that in addition to the absence of extended 
states, the bipolaron (QS) cannot occur with only two sites. 

Setting  $\psi_{i,i}=x$, $\psi_{j,j}=y$ and $\psi_{i,j}=z$, using the normalization 
$x^{2}+y^{2}+2 z^{2}=1$, (\ref{Varform}) becomes

\begin{eqnarray}
	F_{v}&=&-\frac{1}{4}-\frac{1}{4} (x^{2}-y^{2})^{2})+ 
	U(x^{2}+y^{2}) \nonumber\\
	 && \quad - \sqrt{2} t (x+y)\sqrt{1-x^{2}-y^{2}}
	\label{dimform}
\end{eqnarray}

Considering as equivalent the extrema obtained by symmetries ($x \rightarrow -x$, 
$y \rightarrow -y$) and ($x \rightarrow y$, $y \rightarrow x$), there 
are up to 4 kinds of extrema to this variational form at $t=0$:
\begin{itemize}
	\item  Bipolaron (S0) with energy $U-1/2$ is a local minimum for $U<1/2$ and becomes 
a saddle point  with only one unstable direction  for $U>1/2$. There are two symmetric such 
solutions located either at site $i$ or at site $j$.
	\item Bipolaron (S1) with energy $-1/4$ is a local minimum for $U>0$ and becomes a maximum 
(with two unstable directions) for $U<0$.
    \item  Bipolaron (2S0) is a non connected state consisting of
$\psi_{i,i}=\psi_{j,j}=1/\sqrt{2}$ and $\psi_{i,j}=\psi_{j,i}=0$. It 
is a maximum (two unstable directions) for $U>0$ and a saddle point (one 
unstable direction) for $U<0$.  
\item  When $0<U<1/2$, there is another extremum which is the 1 star sister (S1/S0) 
described in appendix (\ref{AppA}). It is a saddle point with one unstable direction. It bifurcates
with bipolaron (S1) at $U=0$ and with bipolaron (S0) at $U=1/2$. There 
are two symmetric such solutions located at site $i$ or $j$.
\end{itemize}

The minimax corresponding to the PN energy barrier for moving 
either bipolaron (S0) or (S1)  is nothing but the unique saddle point
which could be (2S0) ($U<0$), (S1/S0) ($0<U<1/2$) or (S1) ($1/2<U$).

\begin{figure} \begin{center}
\includegraphics{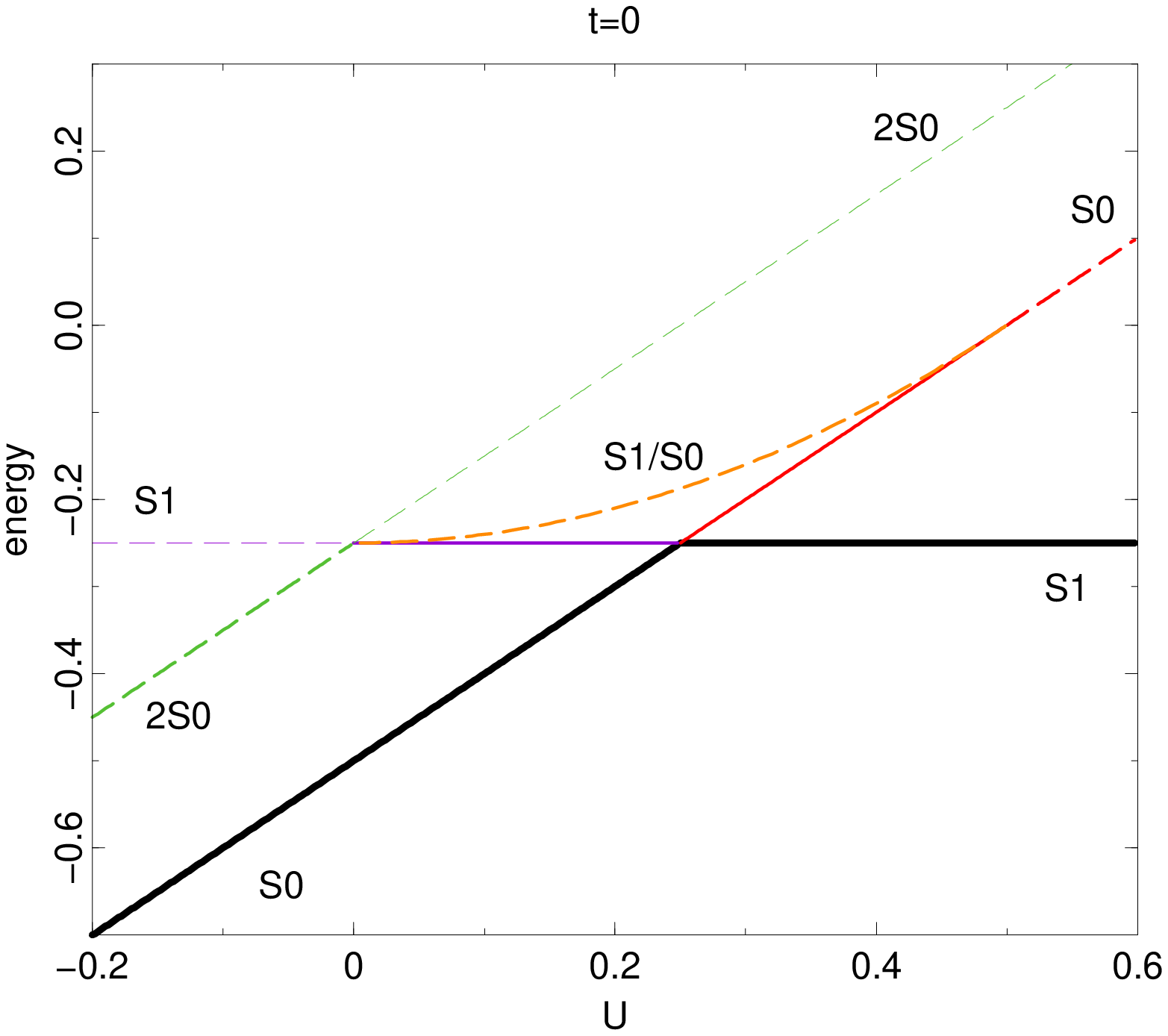}
\includegraphics{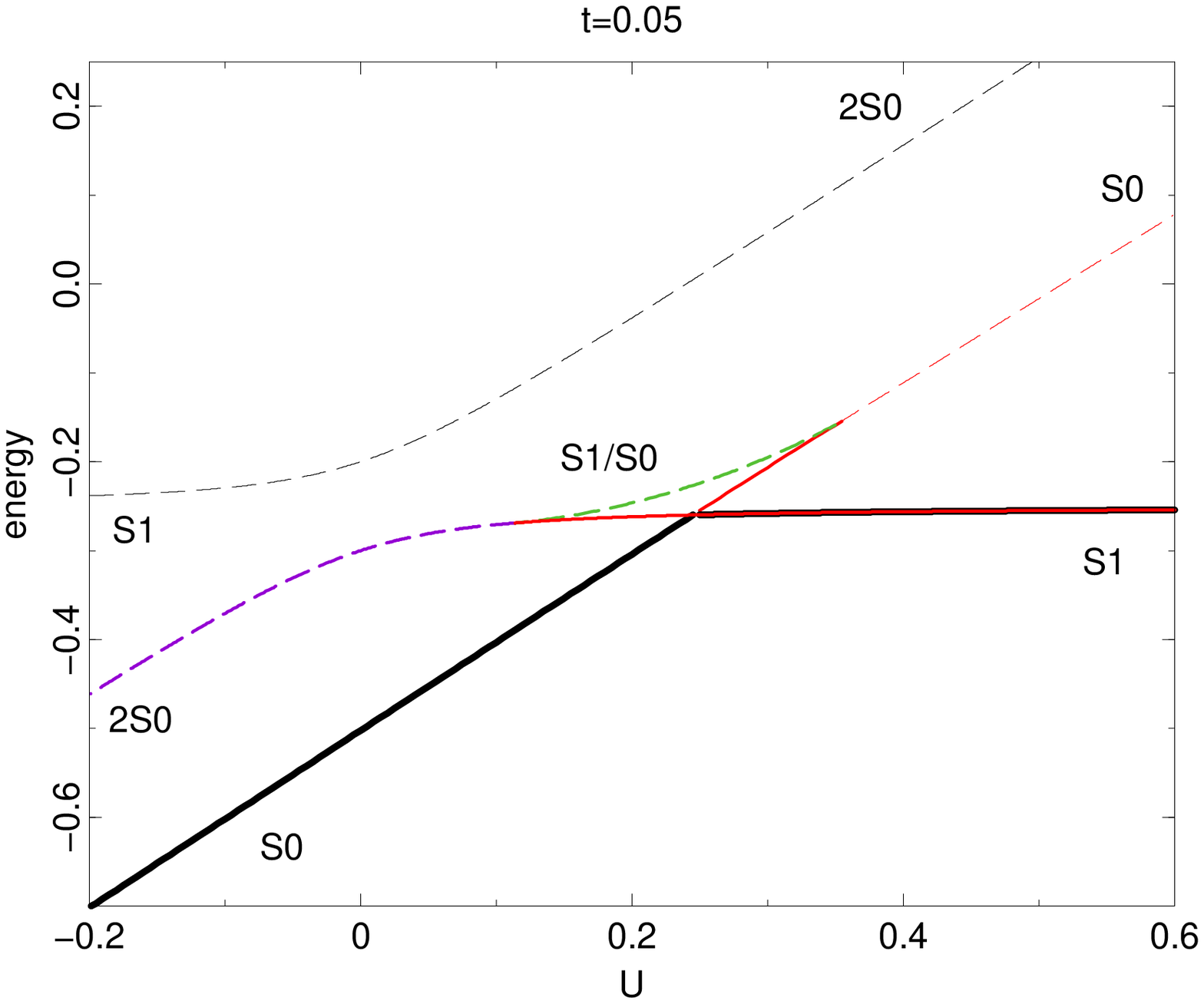}
\caption{\label{fig:14} Bipolaron energies versus $U$ at the 
anti-integrable limit $t=0$ (top) and at $t=0.05$ (bottom).
Ground-state (thick full line), stable bipolaron (full line), minimax 
(thick dashed line), maximum (thin dashed lines) }
\end{center}\end{figure}

At $t=0$ and $U=0$,  bipolarons (S1) and (2S0), which are both spatially 
symmetric, have also the same energy and the same electronic
density (see fig.\ref{fig:14}). When $t\neq 0$, this degeneracy is 
raised as shown in fig.\ref{fig:14}.

\section{Quantum Symmetries of Bipolarons} \label{AppC}

Eventhough no nontrivial quantum symmetry appears for the bipolaronic ground-states 
of our model, it is worthwhile to going forward  now some further work of ours  and discuss 
the possibility of nontrivial quantum symmetries.  Actually, such symmetries are
already \textit{latent} in the anti-integrable limit and could be 
manifested easily in appropriately modified models. 

As we pointed out, in the anti-integrable limit, only the modulus of 
$\psi_{m,n}$ is determined but not the phases. This degeneracy is 
expected to be lifted by the perturbation from this limit due the electronic kinetic 
energy. However, it might not be completely lifted in some cases. 

This situation may occur for bipolarons associated with a
lattice distortion (or equivalently an electronic density) which has the square  symmetry
of the lattice (group $C_{4v}$). This symmetry group has only two 
generators, which 
are for example the $\pi/2$ rotation and the reflection with respect to the $x$ axis.
Any of the symmetry transformations change only  the phase of the electronic wave
function $\{\psi_{m,n}\}$ but not its modulus $\left|\psi_{m,n}\right|$.

\begin{figure} \begin{center}
\includegraphics{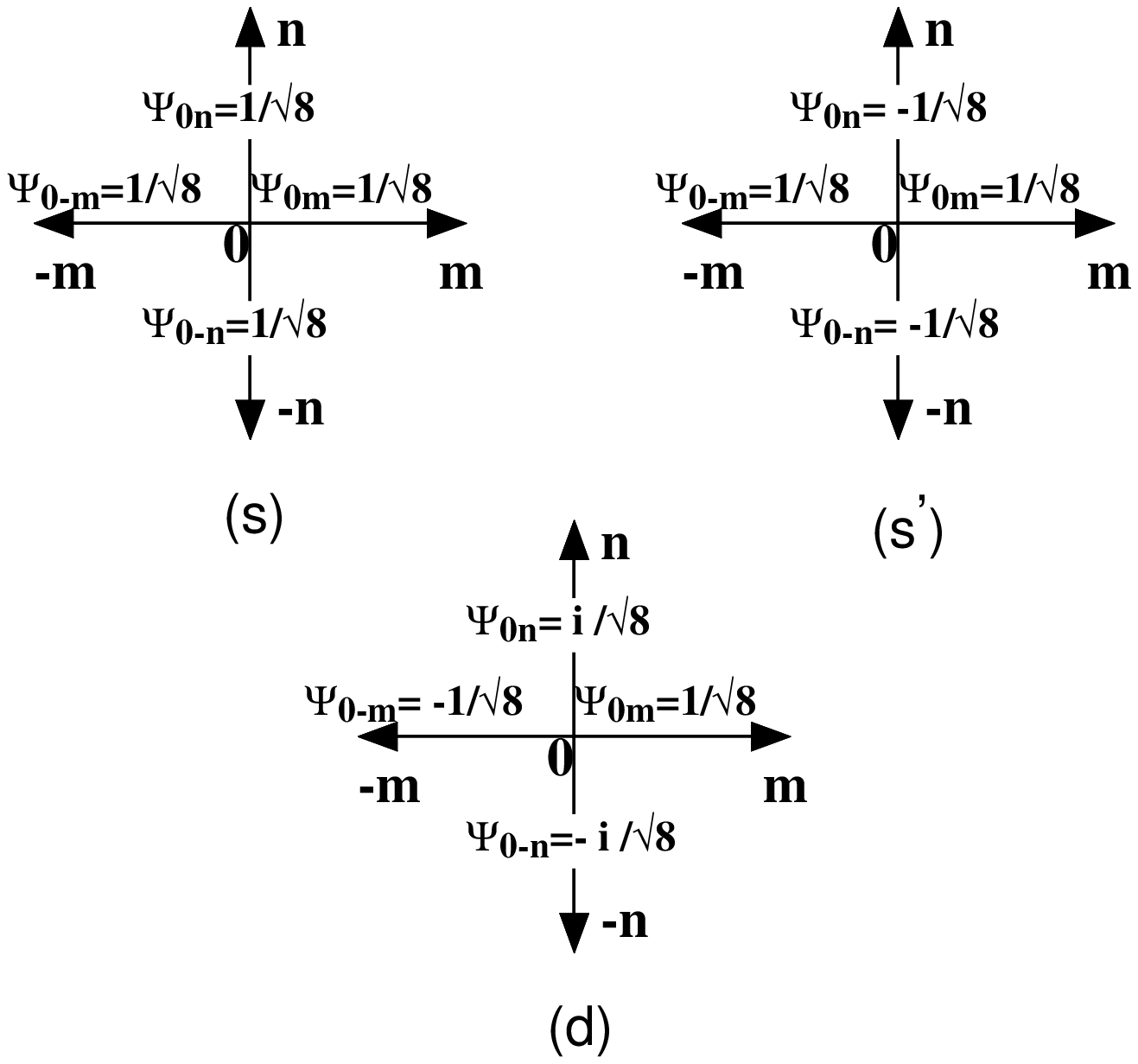}
\caption{\label{fig:15}  Three (QS) bipolarons at the anti-integrable 
limit with the different quantum symmetries $(s)$ (top left),$(s^{\prime})$ (top 
right) and  $(d)$ (bottom) .}
\end{center}\end{figure}

There are three possible group representations for $C_{4v}$ usually denoted A, B and E
in textbooks of crystallography \cite{PM70}. We denote them $(s)$, $(s^{\prime})$ and  
$(d)$ respectively
\footnote{In principle, the $(d)$ symmetry characterizes the 
dimension $5$ representation $l=2$ of the continuous rotation group $O(3)$ in three 
dimensions. We still use this terminology for the symmetry group of the square 
lattice although it is not appropriate, but because it is the most 
standard one in symmetry theory of superconductivity.}).
(see fig.\ref{fig:15})
\begin{itemize}
	\item  When $\{\psi_{m,n}\}$ has the $(s)$ symmetry, it is unchanged by any
	symmetry  operation. This representation has obviously dimension $1$.

	\item   When $\{\psi_{m,n}\}$ has the $(s^{\prime})$ symmetry, 
	$\{\psi_{m,n}\}$ is changed into $\{-\psi_{m,n}\}$, by a $\pi/2$ rotation of the lattice.
It is unchanged  by reflection with respect to the $x$ axis. The 
other transformations are obtained by combinations of these ones.
This representation has also dimension $1$.

	\item   For the $(d)$ symmetry, $\{\psi_{m,n}\}$ is changed into 
$\{i\psi_{m,n}\}$, by a $\pi/2$ rotation of the lattice and  
$\{\psi_{m,n}\}$ is changed into  $\{\psi_{m,n}^{*}\}$, by reflection with respect to 
the $x$ axis. This representation has dimension $2$.
\end{itemize}

We now note that, in the anti-integrable limit, bipolaron (S0) always has the 
symmetry $(s)$. For bipolaron (S1), which does not have the square symmetry but only an 
axis of symmetry, the symmetry is too poor to generate a $d$ 
symmetry. Bipolaron (QS) is more interesting because it has the square 
symmetry for its electronic density but its quantum wave function may 
have three different quantum symmetries  $(s)$,  $(s^{\prime})$ and  $(d)$ respectively 
(see fig\ref{fig:15}). These three states are degenerate in the 
anti-integrable limit but the electronic kinetic energy lifts this degeneracy.
For a Laplacian-like kinetic energy as in the model treated where $t>0$ in (\ref{hamiltelectr}), the 
prefered quantum symmetry is $(s)$. However, symmetry $(d)$ can be easily favored 
when there are next nearest-neighbor transfer integrals with negative 
signs in the electronic kinetic energy.

In summary, we described in appendices (\ref{AppA}) and (\ref{AppC}) 
a systematic method that allows one
to construct all the possible bipolaron solutions existing at the anti-integrable limit $t=0$.
There are bipolaron states with no continuous degeneracy and others with a
continuous degeneracy. Bipolarons (S0) and star multisinglets with 
$N_{2}$ small appear to be the best candidates for bipolaronic 
ground-states when the electronic kinetic energy is switched to be 
non-zero. Star sisters may appear as minimaxes but are not found as 
ground-states. There are bipolarons with the
square lattice symmetry (for example (QS)) which may have nontrivial quantum symmetries $(s^{\prime})$
or $(d)$ instead of $(s)$.


\begin{thebibliography}{99}
\bibitem{BCS57} 
J. Bardeen, L.N. Cooper and J.R. Schrieffer,
Phys.Rev. \textbf{106 } (1957) 162--164 and
\textbf{108} (1957) 1175--1204

\bibitem{Mig58} A.B. Migdal, Zh. Eksperim. Fiz. \textbf{34} (1958) pp. 1438 and
Soviet Phys.-JETP \textbf{7} (1958) pp. 996

\bibitem{McM68}
W.L. McMillan, Phys. Rev. \textbf{167} (1968) pp. 331 

\bibitem{BM86} J.G. Bednorz and K.A. M\"uller, Z.Phys. \textbf{B64} (1986) pp. 1796

\bibitem{MB93} K.A. M\"uller and G. Benedek, \textit{Phase Separations in Cuprate Superconductors}
World Scientific Pub. (the Science and Culture series) (1993)

\bibitem{Wal96} J.R. Waldram \textit{Superconductivity of Metals and Cuprates}
IOP Publishing Ltd (1996)

\bibitem{ASL95} A.S. Alexandrov, E.K.H. Salje and W.H. Liang, \textit{Polarons and
Bipolarons in High Tc Superconductors and related Materials} (1995)
Cambridge University Press

\bibitem{Lan33} L. Landau, Phys. Z. Sowjetunion \textbf{3} (1933) 664

\bibitem{AAR86} A.S. Alexandrov, J. Ranninger and S. Robaszkiewicz,
Phys.Rev. \textbf{B33} (1986) 4526--4552

\bibitem{EH76} D. Emin and T. Holstein, Phys. Rev. Lett. \textbf{36} (1976) 4526; D. Emin, Phys.Today 
(june 1982) pp. 34

\bibitem{CRF98} B.K. Chakraverty, J. Ranninger and D. Feinberg, Phys. Rev. Lett. \textbf{81}
(1998) 433--436

\bibitem{AQ89} S. Aubry and P. Quemerais, in \textit{Low Dimensional Electronic Properties
of Molybdenum Bronzes and Oxides} Ed. Claire Schlenker, Kluwer Academic Publishers
Group (1989)  295--405 

\bibitem{AAR92} S. Aubry, G. Abramovici and J.L. Raimbault, J. Stat. Phys.
\textbf{67} (1992) 675--780

\bibitem{Aub93} S. Aubry, J.Physique IV colloque (Paris)  \textbf{C2} \textbf{3} (1993) 349--355

\bibitem{Aub93b} S. Aubry in ref.\cite{MB93},  304--334
\bibitem{Aub95b} S. Aubry in ref.\cite{ASL95}, 271--308

\bibitem{PA99} L. Proville and S. Aubry, in preparation

\bibitem{Pro98} L. Proville \textit{Structures Polaroniques et Bipolaroniques
dans le Model\`ele de Holstein Hubbard Adiabatique a deux electrons et ses Extensions}
PhD Dissertation, University Paris XI, (1998)

\bibitem{PA98} L. Proville and S. Aubry, Physica \textbf{113D} (1998) 307--317

\bibitem{Aub95} S. Aubry, Physica \textbf{86D} (1995) 284--296

\bibitem{KAT98} G. Kalosakas, S. Aubry and G. Tsironis, Phys. Rev. \textbf{B58} (1998) 3094--3104

\bibitem{LeD83} P.Y. Le Da\"eron, \textit{Transition Metal-Isolant dans les chaines de Peierls}
PhD Dissertation, University Paris XI, (1983)

\bibitem{MA96} J.L. Mar\'{\i}n and S. Aubry,  Nonlinearity \textbf{9} (1997) 1501--1528 

\bibitem{BMK94} C. Baesens and R.S. MacKay, Nonlinearity \textbf{7} (1994) 
59--84

\bibitem{MA94} R.S. MacKay and S. Aubry,  Nonlinearity \textbf{7} (1994) 
1623--43

\bibitem{And87} P.W. Anderson, G. Baskaran, Z. Zou and T. Hsu,
Phys.Rev.Lett. \textbf{58} (1987) 2790--2793

 \bibitem{AC98} S. Aubry and T. Cretegny, Physica \textbf{119D} (1998) 34--46
 
\bibitem{AC99}T. Cretegny, PhD Dissertation (ENS Lyon 1998) and
T. Cretegny and S. Aubry, in preparation

\bibitem{Pin97} D. Pines, Physica \textbf{282C} (1997)  273--278 


\bibitem{BMc98} C. Baesens and R. MacKay, \textit{Excited states in the Adiabatic Holstein Model}
, J. Phys. A, in press (1998)

\bibitem{PM70} \textit{Spectres de vibration et sym\'etrie des cristaux} 
 H. Poulet et J-P. Mathieu, Gordon \& Breach Paris (1970)

\end{thebibliography}
\end{document}